\documentclass[useAMS,usenatbib]{mn2e}

\usepackage{threeparttable}
\usepackage{graphics,graphicx,epsfig,ulem}
\usepackage{multirow}
\usepackage{amsmath}
\usepackage[hyphens]{url} 
\usepackage{booktabs} 
\usepackage{marvosym} 
\usepackage{xtab} 
\usepackage{caption} 
\usepackage[T1]{fontenc} 
\usepackage{xcolor} 
\usepackage{soul} 
\def\la{\mathrel{\hbox{\rlap{\hbox{\lower4pt\hbox{$\sim$}}}{\raise2pt\hbox{$<$}}}}}
\def\ga{\mathrel{\hbox{\rlap{\hbox{\lower4pt\hbox{$\sim$}}}{\raise2pt\hbox{$>$}}}}}

\newcommand{\hoix}{Holmberg IX X-1 }
\newcommand{\xmm}{\textit{XMM-Newton} }
\newcommand{\swift}{\textit{Swift} }
\newcommand{\nustar}{\textit{NuSTAR} }
\newcommand{\comptt}{{\sc comptt} }

\title[X-ray spectral evolution of Ho IX X-1]{The X-ray spectral evolution of the ultraluminous X-ray source Holmberg IX X-1}
\author[W. Luangtip, T. P. Roberts and C. Done]{Wasutep Luangtip,$^{1,2}$\thanks{E-mail:wasutep@g.swu.ac.th} Timothy P. Roberts$^{1}$ and Chris Done$^{1}$\\
$^{1}$Centre for Extragalactic Astronomy, Department of Physics, University of Durham, South Road, Durham, DH1 3LE, UK\\
$^{2}$Department of Physics, Faculty of Science, Srinakharinwirot University, 114 Sukhumvit 23, Wattana, Bangkok 10110, Thailand
}
\begin{document}

\date{Draft, \today}

\pagerange{\pageref{firstpage}--\pageref{lastpage}} \pubyear{2015}

\maketitle

\label{firstpage}

\begin{abstract}
We present a new analysis of X-ray spectra of the archetypal ultraluminous X-ray source (ULX) Holmberg IX X-1 obtained by the {\it Swift}, {\it XMM-Newton} and {\it NuSTAR} observatories.  This ULX is a persistent source, with a typical luminosity of $\sim$10$^{40}$ erg s$^{-1}$, that varied by a factor of 4 -- 5 over eight years.  We find that its spectra tend to evolve from relatively flat or two-component spectra in the medium energy band (1-6 keV), at lower luminosities, to a spectrum that is distinctly curved and disc-like at the highest luminosities, with the peak energy in the curved spectrum tending to decrease with increased luminosity.  We argue that the spectral evolution of the ULX can be explained by super-Eddington accretion models, where in this case we view the ULX down the evacuated funnel along its rotation axis, bounded by its massive radiatively driven wind. The spectral changes then originate in enhanced geometric beaming as the accretion rate increases and wind funnel narrows, causing the scattered flux from the central regions of the supercritical flow to brighten faster than the isotropic thermal emission from the wind, and so the curved hard spectral component to dominate at the highest luminosities.  The wind also Compton down-scatters photons at the edge of the funnel, resulting in the peak energy of the spectrum decreasing.  We also confirm that \hoix displays spectral degeneracy with luminosity, and suggest that the observed differences are naturally explained by precession of the black hole rotation axis for the suggested wind geometry.
\end{abstract}


\section{Introduction}

Ultraluminous X-ray sources (ULXs) are extra-galactic, non-nuclear point sources, with X-ray luminosities $>$ 10$^{39 }$ erg s$^{-1}$, equivalent to or in excess of the Eddington limit for 10 $M_{\rm \odot}$ black holes (BHs; see \citealt{feng2011,roberts2007} for reviews).  Given their non-nuclear locations, they cannot be powered by accretion onto the central supermassive black hole of their host galaxy.  However, due primarily to their extreme luminosities, it has been proposed that ULXs might be a hitherto missing class of massive black holes, the  intermediate mass black holes (IMBHs), with 10$^{2}$ $<$ $M_{\rm BH}$ $<$ 10$^{5}$ (e.g. \citealt{colbert1999,miller2003}), that accrete material at the sub-Eddington rates we typically see in Galactic black hole binaries (BHBs).  Indeed, some of the most luminous ULXs remain good candidates for sub-Eddington accretion onto IMBHs, for instance HLX-1 in ESO 243-49 \citep{farrell2009} and a sample of the most extreme ULXs discussed by \citet{sutton2012}, including NGC 2276 3c \citep{mezcua2015}.  However, recent mass constraints on two ULXs  \citep{liu2013,motch2014} suggest the presence of stellar mass BHs (sMBHs) in these objects, that must be accreting at super-Eddington rates.  These results, combined with differences in both the X-ray spectral and timing characteristics of many ULXs when compared to sub-Eddington BHBs (e.g. \citealt{gladstone2009,sutton2013}), indicate that this could be the nature of most ULXs, although an intriguing possibility is that some ULXs may harbour larger stellar remnant BHs (MsBHs, 20 $<$ $M_{\rm BH}$ $<$ 100) that may form in metal-poor environments (e.g. \citealt{zampieri2009}).  Similarly, we now know that at least one ULX harbours an extreme super-Eddington neutron star, given the detection of pulsations from an object in M82 \citep{bachetti2014}.

The X-ray spectra of ULXs have been widely studied since the launch of the \xmm and \textit{Chandra} X-ray observatories, and so knowledge of their typical properties has grown substantially in the last decade.  It has been shown that the X-ray spectra of ULXs in the 0.3-10 keV band are composed of two individual components: one dominating at energies below 1 keV (the soft excess) and another dominating at photon energies above 2 keV (the hard component; e.g. \citealt{miller2003}).  The very cool temperatures ($\la$ 0.2 keV) obtained when fitting the soft excess with accretion disc models were initially used to imply the existence of IMBHs in many bright ULXs (e.g. \citeauthor{miller2004b} 2004b).  However, subsequent studies of the highest quality X-ray spectra of ULXs \citep{stobbart2006,gladstone2009} demonstrated that the hard component turns over at photon energies $\sim 3 - 7$ keV, a result that has more recently been confirmed by broad band \nustar observations that show this break extends well above 10 keV \citep{bachetti2013,walton2014}.  This spectrum is not generally observed in Galactic BHBs, which when combined with the high luminosities of ULXs implies it is likely to represent a new mode of accretion -- putatively called the \textit{ultraluminous state} -- in which sMBHs are accreting material at super-Eddington rates \citep{gladstone2009}.

A physical scenario that could describe super-Eddington accretion processes has been discussed by several theoretical and simulation-based works (see e.g. \citealt{poutanen2007,kawashima2012,takeuchi2013,takeuchi2014,sadowski2015b})\footnote{Although other models may not be discounted based on current data, for example locally inhomogeneous discs \citep{miller2014}.}.   At super-Eddington rates accreting BHs are predicted to increase their disc scale height (H/R) to $\sim$ unity as the disc interior becomes advection-dominated and so heats up; the outer layers of the disc then become loosely bound, and the intense radiation release from the disc launches a massive, outflowing wind within a photospheric radius, that forms a funnel-like structure around the rotational axis of the BH.   Observational support for this scenario has been gradually emerging over the last few years. \citet{sutton2013} showed that the good quality spectra of ULXs observed by \xmm predominantly appear like broadened discs at luminosities below $\sim$3 $\times$ 10$^{39 }$ erg s$^{-1}$, where a sMBH would be accreting material at $\sim$ the Eddington rate; above this luminosity, the ULXs accrete material at rates that may substantially exceed Eddington, and the spectra become two component, as described above.  The harder component is well-modelled by an optically thick Comptonising corona ($kT_{\rm e} \sim 2$ keV, $\tau \sim 10$; see e.g. \citealt{pintore2012,vierdayanti2010}); however recent works have interpreted this spectrum as more likely to originate in emission from the hot, inner parts of the accretion flow (e.g. \citealt{middleton2011,kajava2012}).  The soft, disc-like component is then interpreted as emission from the optically thick and massive outflowing disc wind (e.g. \citealt{kajava2009}, although see \citealt{miller2013}).  

This interpretation is strongly supported by work considering the spectral and temporal variability properties of ULXs, where very high levels of flux variability on timescales of $\la 100$s of seconds are associated with the hard spectral component when the spectrum is dominated by the soft component (i.e. the ULX is in a soft ultraluminous, or SUL, regime; \citealt{sutton2013}).  This can be explained by the wind being an inhomogeneous medium, where optically-thick clumps pass through the observer's line-of-sight to the central regions of the accretion flow and scatter away the hard photons \citep{middleton2011,sutton2013,middleton2015}.  That this variability is seen when the spectrum is in the SUL regime supports the notion that both accretion rate and viewing angle dictate the observational appearance of ULXs; in the SUL regime the line-of-sight intersects the wind, which diminishes the hard component, and adds extrinsic variability.  In contrast, when viewed down the funnel, a hard ultraluminous (HUL) spectrum is observed, with little or no strong variability.  The major missing piece of the puzzle in this model is a direct detection of the wind material, from absorption or emission lines in the optically thin phase of the wind that must be present as the wind diffuses away from the disc; but this may be the origin of residuals in the soft X-ray spectra of ULXs \citep{middleton2014}, and broad emission lines seen in the optical spectra of ULX counterparts \citep{fabrika2015}.

\hoix is a nearby ULX ($d$ = 3.42 Mpc; \citealt{liu2005}) located close to the dwarf galaxy Holmberg~IX.  It is a persistent source with luminosity $\sim$10$^{40}$ erg s$^{-1}$ that displays variations in flux by a factor of 3 -- 4 on timescales of days \citep{kaaret2009,kong2010}.  The source was first discovered by the \textit{Einstein} Observatory \citep{fabbiano1988} and has been well studied over the past thirty years due to its proximity and so high X-ray flux.  Despite some initial uncertainty in its nature, with some discussion as to whether it was a background QSO or a supernova remnant, it was confirmed as a likely ULX at the start of the \xmm and {\it Chandra} epoch \citep{la_paprola2001}.   Subsequent studies focusing on the high quality spectra obtained by \xmm have shown that they are typical of a ULX in the HUL regime (e.g. \citealt{stobbart2006,sutton2013}), and a study of its X-ray spectral variability by \citet{vierdayanti2010} showed that the variability was not a simple function of luminosity, but that subtle variations occurred that appeared independent of accretion rate.  A more recent study incorporating broad band (0.3-30 keV) spectra from \nustar demonstrated that the spectra can be explained by two optically-thick thermal components, similar to the spectra seen in the 0.3-10 keV range; furthermore the spectral evolution was explored, with its characteristics hypothesised to be indicative of physical changes in the highly dynamical, outflowing wind or the evolution of the hot inner region of the disc itself \citep{walton2014}.  Given its high flux and hard spectrum, \hoix was a natural choice to search for absorption/emission features in the Fe K band that may be indicative of material in an outflowing wind.  However, none were found to stringent limits by \citeauthor{walton2013a} (2013a); this is perhaps unsurprising, though, given the HUL spectra that are indicative of \hoix being viewed down the funnel in its disc/wind structure, rather than through its wind.

Although we now have a provisional working model for the super-Eddington processes that occur in ULXs, much of the detail remains to be determined.  One obvious way of better assessing the physical mechanisms and/or geometry at play in ULXs is through the detailed study of bright, individual archetypes for the class.  In this work, we analyse the variation in spectra of \hoix using data obtained from the {\it Swift}, \xmm and \nustar X-ray observatories, in order to constrain the spectral evolution of the ULX with luminosity, and so better determine its physics.  The paper is laid out as follows. In Section~\ref{sec:observations and data reduction}, we explain how we select the X-ray data, how we reduce it and how we create the ULX spectra.  The details of the spectral analysis are presented in Section~\ref{sec:spectral analysis and results} and we discuss what we learn from the variability of the spectra in Section~\ref{sec:discussion}. Our findings are summarised in Section~\ref{sec:conclusion}.

\section{Observations and data reduction}
\label{sec:observations and data reduction}

\subsection{\textit{Swift} data}

The X-ray variability of  \hoix has been monitored intermittently by \swift during the last decade, some observations of which were part of a monitoring programme for bright ULXs (e.g. \citealt{kaaret2009}).  We searched for useful \textit{Swift} observations of \hoix  in the \textit{Swift} data archive catalogue\footnote{\url{http://swift.gsfc.nasa.gov/archive/}}, using a cone search with radius of 11 arcminutes such that the ULX position is always located within the field of view of the \swift X-ray telescope (XRT).  Only the data obtained in photon counting (PC) mode were selected in order to obtain 2D images of the observations.  We also selected only those observations in which the XRT exposure time is $>$ 40 seconds, in order to (typically) obtain at least 10 photon counts from the ULX in each observation.  After this step, we ended up with 514 useful observations: 132 observations in which the ULX position lies within a circular region of 5 arcminute radius centred on the XRT detector aim point (hereafter on-axis observations); and a further 382 observations that are \swift monitoring observations of the nearby galaxy M81, in which the ULX position is $>$ 5 arcminutes from the XRT detector aim point  (hereafter off-axis observations) but still lies with in the field of view of the XRT. 

We reduced all selected observations of \hoix following the \swift XRT Data Reduction Guide version 1.2\footnote{\url{http://swift.gsfc.nasa.gov/analysis/xrt_swguide_v1_2.pdf}}.  In brief, the raw data were reduced using the script {\sc xrtpipeline}. Using the default parameter values provided by the script, bad pixels were removed and clean event files were created from the good grade events (grade 0 -- 12). Then, source and background spectra were extracted from the clean event files using the script {\sc xrtproducts}, which also automatically created the appropriate auxiliary response files (ARFs) and response matrix files (RMFs).  In all cases, the source spectra were extracted from a circular region of 47 arcseconds radius located at the source position, corresponding to the 90\% encircled energy radius at 1.5 keV for the on-axis point spread function (PSF) of the \swift XRT\footnote{\url{http://swift.gsfc.nasa.gov/analysis/threads/uvot_thread_spectra.html}; note that this decreases to 63\% encircled energy for the same radius at 11 arcminutes off-axis.}.  For the background spectra, the extraction was divided into two cases; for the on-axis observations, the background spectra were extracted from an annular region (of inner and outer radii 75 and 150 arcseconds respectively) around the source extraction area; in the case of off-axis observations, we extracted the background spectra from a source free, circular region of radius 150 arcseconds next to the source extraction aperture.  The properties of the spectra obtained from the \swift observations are listed in Table~A1.

\subsection{\xmm data}
\label{sec:xmm data}

\hoix has been observed several times over the last decade-and-a-half by \textit{XMM-Newton}; these observations have been analysed by many authors, as shown in column 6 of Table~\ref{tab:xmm_obs}.  Here we include all this \xmm data in our analysis.  We began by searching for observations of the ULX in the \xmm data archive\footnote{\url{http://nxsa.esac.esa.int}}; all 15 observations of \hoix that we found are tabulated in Table~\ref{tab:xmm_obs}. We reduced the \xmm data using \xmm Science Analysis Software ({\sc sas})\footnote{\url{http://xmm.esac.esa.int/sas/}} version 13.5.0, following the instructions in the {\sc sas} thread web pages\footnote{\url{http://xmm.esac.esa.int/sas/current/documentation/threads/}}. We reprocessed the observation data files (ODFs) to obtain new calibrated and reprocessed event files, using the scripts {\sc epproc} and {\sc emproc} for the PN and MOS data respectively.  Then, we filtered the event files for background particle flaring events using the {\sc sas} task {\sc espfilt}, which generated the clean PN and MOS event files.

We extracted PN and MOS spectra from the good grade events suggested by the {\sc sas} thread, i.e. {\sc flag} = 0 and {\sc pattern} $\leq 4$ for the PN and 12 for the MOS, respectively.  The source spectra were extracted from a circular aperture of 50 arcsecond radius around the source position, corresponding to $\sim$90\% encircled energy at 1.5 keV of the PN and MOS on-axis PSFs\footnote{\url{http://xmm.esac.esa.int/external/xmm_user_support/documentation/uhb_2.1/node17.html}; this decreases to 43\% encircled energy for the same radius at the edge of the detectors.}.  The background spectra were extracted from a source-free, circular area of 80 arcsecond radius near to the ULX, with the same off-axis angle to that of the source and, if possible, from the same CCD as the ULX.  The corresponding RMFs and ARFs were generated with the {\sc sas} scripts {\sc rmfgen} and {\sc arfgen}.  The spectra were then grouped to have a minimum of 20 counts per bin to utilise the $\chi^{2}$ optimisation method during spectral modelling.  The total number of counts obtained from each \xmm observation is shown in column 4 of Table~\ref{tab:xmm_obs}. 

In the broadband spectral analysis of the source (Section~\ref{sec:xmm_nustar spectral analysis}), we prepared the \xmm spectra using the same method as that of the previous broadband study of the ULX \citep{walton2014}. Hence we segregated the \xmm spectra that were observed simultaneously with the \nustar observatory in October and November 2012 into epoch 1 and epoch 2 (See Table~\ref{tab:xmm_obs}).  The spectra obtained from the same detectors in each epoch were stacked together using the {\sc ftool addspec}\footnote{\url{http://heasarc.gsfc.nasa.gov/ftools/caldb/help/addspec.txt}}, in order to create a single spectrum of  PN, MOS1 and MOS2 for each observational epoch.  The spectra were then grouped to have a minimum of 50 counts per bin.

    \begin{table*}
      \centering
      \caption{\xmm observations of \hoix}\label{tab:xmm_obs}
      \smallskip
      \begin{threeparttable}
          \begin{tabular}{lccccc}
\hline
ObsID & Obs. Date & PN/M1/M2 Exp.$^{a}$ & Total Counts$^{b}$ & $\theta$$^{c}$ & References$^{d}$ \\          
 &  &  (ks) & & (arcmin) & \\
\hline          
0111800101$^{e}$	&    2001-04-22		&    --/76.0/76.7	&    	65072	&    	10.70 & 1,2,3,4,5,6,7,8 \\
0111800301$^{f}$	&    2001-04-22		&    -		&    -		&    10.70 &  -- \\
0112521001	&    2002-04-10		&    5.9/9.5/9.9	&    	24138	&    	2.09  &  1,2,3,6,8,9,10,11,12,13,16\\
0112521101	&    2002-04-16		&    7.2/8.6/8.5	&    	29409	&    	2.09  & 2,3,6,9,10,11,12,13,14,15,16 \\
0200980101	&    2004-09-26	&    	38.0/62.2/65.0	&    	125946	&    	2.18 &   1,2,3,4,5,6,7,9,10,11,12,16,17,18,19,20,21\\
0657801601$^{f}$	&    2011-04-17	&    	--	&    	2634	&    	4.50  &  11,22,23\\
0657801801	&    2011-09-26	&    	3.6/10.8/13.3	&    	28676		&    4.50  &  2,11,16,22,23\\
0657802001	&    2011-03-24	&    	2.5/4.7/4.6	&    	7154		&    4.50  &  2,11,16,22,23\\
0657802201	&    2011-11-23		&    11.7/15.3/15.2	&    	60861		&    4.50  &  2,11,16,22\\
0693850801$^{ep1}$	&    2012-10-23	&    	5.7/7.8/8.0	&    	24553		&    2.17 &   24\\
0693850901$^{ep1}$	&    2012-10-25		&    4.9/8.3/11.1	&    	28040		&    2.17  &  24\\
0693851001$^{ep1}$	&    2012-10-27		&    3.9/5.3/6.5	&    	21282		&    2.17  &  24\\
0693851101$^{ep2}$	&    2012-11-16		&    2.5/4.4/4.4	&    	24115		&    2.17  &  24\\
0693851701$^{ep2}$	&    2012-11-12		&    6.1/9.3/9.0	&    	50831		&    2.17 &   24\\
0693851801$^{ep2}$	&    2012-11-14		&    6.6/8.5/8.5	&    	51356		&    2.17  &  24\\
\hline

         \end{tabular}
         \begin{tablenotes}
         \item \textbf{Note.} $^{a}$The good exposure times obtained from the PN, MOS1 and MOS2 detectors. $^{b}$The total useful counts from the source, observed during the good exposure times, from a combination all three EPIC detectors.  $^{c}$The off-axis angle of the source (measured from the EPIC aim point). $^{d}$References to previous studies of this ULX using the pertinent data.  These are: (1) \citet{winter2007}; (2) \citet{sutton2014}; (3) \citet{miller2013}; (4) \citet{gonzalez-mart2011}; (5) \citet{gladstone2009}; (6) \citet{sutton2013}; (7) \citet{caballero-garc2010}; (8) \citet{winter2006}; (9) \citet{walton2012}; (10) \citet{kajava2009}; (11) \citet{pintore2014}; (12) \citet{vierdayanti2010}; (13) \citet{wang2004}; (14) \citet{poutanen2007}; (15) \citet{stobbart2006}; (16) \citet{middleton2015}; (17) \citet{heli2009}; (18) \citet{hui2008}; (19) \citet{grise2011}; (20) \citet{dewangan2006}; (21) \citet{berghea2013}; (22) \citet{sazonov2014}; (23) \citet{dewangan2013}; (24) \citet{walton2014}.  $^{e}$Only MOS spectra are extracted from the observation as the source position falls outside of the PN detector area. $^{f}$We exclude the observation from the analysis as the good exposure time is too low to obtain good quality spectra. $^{ep1}$The spectra extracted from these observations were stacked together and analysed as epoch 1 in the broadband spectral analysis. $^{ep2}$The spectra extracted from these observations were stacked together and analysed as epoch 2 in the broadband spectral analysis.
         \end{tablenotes}
      \end{threeparttable}
    \end{table*}

\subsection{\nustar data}
\label{sec:nustar data}

\hoix was one of the ULX targets to be observed by \nustar during the primary mission phase; it was observed simultaneously with \xmm in October and November 2012 (see Table~\ref{tab:nustar_obs}; \citealt{walton2014}). In this work, these public data are also included in our analysis.  We reprocessed and reduced the data using the \nustar data analysis software ({\sc nustardas}) version 1.4.1, which is part of the {\sc heasoft} software version 6.16\footnote{\url{http://heasarc.gsfc.nasa.gov/docs/software/lheasoft/}}, using the \nustar instrument calibration files  (\nustar CALDB) version 20140414.  We extracted the spectra from this data using the wrapper task {\sc nupipeline}\footnote{\url{https://heasarc.gsfc.nasa.gov/ftools/caldb/help/nupipeline.html}}, which generated the filtered event files and then extracted the source and background spectra as well as appropriate RMFs and ARFs from the cleaned event and calibration files. The source spectra, all of which were on-axis in the detector, were extracted from a circular aperture of 101 arcsecond radius around the source position, corresponding to the 80\% encircled energy radius of the \nustar detector PSF\footnote{\url{http://www.nustar.caltech.edu/uploads/files/nustar_performance_v1.pdf}}. The background spectra were extracted from a source-free annulus (of inner and outer radius 170 and 270 arcseconds respectively) around the source extraction region. 

Similarly to the \xmm spectra, we divided the \nustar spectra into epoch 1 and epoch 2, segregating by the month in which they were observed (see column 5 of Table~\ref{tab:nustar_obs}), following the method used in the previous broadband study of the ULX \citep{walton2014}.  The spectra obtained from the same \nustar detectors in each epoch were stacked together using the script {\sc addspec}, in order to create a single spectrum from the FPMA and FPMB detectors for each observational epoch.  Finally, the spectra were grouped to have a minimum of 50 counts per bin.

    \begin{table}
      \centering
      \caption{\nustar observations of \hoix}\label{tab:nustar_obs}
      \smallskip
      \begin{threeparttable}
          \begin{tabular}{lcccc}
\hline
ObsID & Obs. Date & Exp.$^{a}$ & Cnts$^{b}$  & Epoch$^{c}$\\          
 &  & (ks) & &\\
\hline          
30002033002	&   2012-10-26		&   31.2		&   13616  &1\\
30002033003	&   2012-10-26		&   88.1		&   41967  &1\\
30002033004$^{d}$	&   2012-11-10		&   --		&   -  &\\
30002033005	&   2012-11-11		&   40.8		&   31450  &2\\
30002033006	&   2012-11-11		&   35.2		&   26399  &2\\
30002033007$^{d}$	&   2012-11-14		&   --		&   -  &\\
30002033008	&   2012-11-14		&   14.5		&   11737  &2\\
30002033009$^{d}$	&   2012-11-15		&   --		&   -  &\\
30002033010	&   2012-11-15		&   49.0		&   37202  &2\\
\hline

         \end{tabular}
         \begin{tablenotes}
         \item \textbf{Note.} $^{a}$The good exposure time obtained from each detector.  $^{b}$The total counts in each observation, from combining the FPMA and FPMB data.  $^{c}$The epoch of observation. The spectra within the same epoch are stacked together and analysed simultaneously (see Section~\ref{sec:xmm_nustar spectral analysis} for the detail of analysis).  $^{d}$We excluded the observation from the analysis as the good exposure time is too low to obtain useful spectral data. 
         \end{tablenotes}
      \end{threeparttable}
    \end{table}

\section{Spectral analysis and results}
\label{sec:spectral analysis and results}

The large number of X-ray spectra of Holmberg IX X-1, from multiple X-ray missions, provides us with an excellent opportunity to further examine the X-ray spectral evolution of this ULX with luminosity.  In this paper, we constrain the shape and hence the evolution of the spectra primarily using a two component model composed of a multi-coloured disc blackbody model (MCD; \citealt{mitsuda1984}) and a Comptonising corona \citep{titarchuk1994}, i.e. {\sc diskbb} + {\sc comptt} in {\sc xspec}.   This model, or very similar models, has been used extensively and very successfully to describe ULX X-ray spectra below 10 keV (see e.g. \citealt{gladstone2009,pintore2014,middleton2015}), with the interesting caveat that the coronae appear cool and optically thick in ULXs modelled in this way ($kT_{\rm e} \sim 1 - 2$ keV; $\tau \ga 5$), most unlike the hot, optically thin coronae seen in Galactic BHBs.  We note that there are other empirical models that can produce similar-shaped spectra (see e.g. \citealt{stobbart2006}); we do not discuss them in this Section, but will return to them later in Section 4 where we use them to illuminate our discussion of the spectral variability.  

We simplify the MCD plus Comptonising corona model by tying the seed photon temperature of the \comptt component to the temperature of the MCD component in order to get better constraints on the model-fitting parameters. It is noted elsewhere that this simplification may lead to unrealistic values of the seed photon temperature in ULX fits, as the measured MCD temperature might not reflect the true temperature of the seed photons from the inner disc due to the obscuration of this region by the putative optically thick corona (see e.g. \citealt{gladstone2009}).  However, in this work, we do not attempt to directly interpret the model parameters; instead, we use the model to evaluate the change in the spectral shape, and to attempt to track the relative contributions of the separate soft and hard components of the spectra.  So, in our case, the simplification of the model should not overtly affect the interpretation of the spectral evolution.  

In fitting models to the spectral data, absorption along the line of sight to \hoix is accounted for by adding two multiplicative absorption components ({\sc tbabs} in {\sc xspec}) into the model, using the interstellar abundances reported by \citet{wilm2000}.  The first absorption component is fixed at 4.06 $\times$ 10$^{20}$ cm$^{-2}$, and accounts for the Galactic column density along our line of sight \citep{dickey1990}, whilst the second component represents absorption external to our Galaxy, most likely in the immediate vicinity of the ULX and/or in its host galaxy.  The spectra were modelled using {\sc xspec}\footnote{\url{http://heasarc.nasa.gov/xanadu/xspec/}} version 12.8.2 over the energy band of 0.3-10 keV (unless otherwise specified).  The best fitting parameters are derived using a $\chi^{2}$ minimisation technique, and throughout this paper their errors are quoted using the 90\% confidence interval. We note that the spectral binning changes between different parts of this section (See Section~\ref{sec:observations and data reduction} and Section~\ref{sec:swift spectral analysis}).  In the analysis of \swift and \xmm spectra, we select a minimum counts per bin that is optimised to extract the best constraints from the data available, whilst in the broadband spectral analysis the spectra are binned to have a minimum of 50 counts per bin in order to replicate the analysis from previous work \citep{walton2014}.

\subsection{\textit{Swift} spectral analysis}
\label{sec:swift spectral analysis}

\subsubsection{Individual spectral analysis}
\label{sec:swift individual spectra}

Unfortunately each individual \swift spectrum contains insufficient data for detailed spectral analysis ($\sim$10 -- 1000 counts).  If we are to obtain the high signal to noise (S/N) spectra required for this study, then we must stack \swift spectra with similar properties together.  A previous spectral study of \hoix using \swift data by \citet{vierdayanti2010} segregated the spectra into luminosity bins using the detector count rates of the ULX.  However, this is not appropriate in this work, as we include off-axis \swift spectra, and the observed count rates for a source with a given flux are a function of the detector effective area which varies with off-axis angle \citep{tagliaferri2004}.  For example, the count rate of a source located at 10 arcminutes off-axis from the detector aim point is lower by $\sim$25 per cent than the count rate of the same source located at the detector aim point. Thus, to avoid this issue, we take into account the differences in the detector response by calculating the observed flux of \hoix in each observation, instead of using the detector count rates.  The fluxes were calculated by modelling the individual \swift spectra with an absorbed power-law ({\sc tbabs*tbabs*powerlaw} in {\sc xspec}), with the absorption modelled as described above.

 The 440 spectra in which $\geq$ 100 counts were detected were grouped to have a minimum of 10 counts per bin, and then each was fitted by the absorbed power-law model.  In addition, to include more \swift data in the analysis, we also considered another 55 observations which have between 50 -- 100 counts.   We again grouped the low count spectra to have a minimum of 10 counts per bin, and the spectra were fitted by the absorbed power-law model.  However, as these spectra cannot constrain all three model parameters, we froze two -- the absorption column external to our Galaxy ($N_{\rm H}$) and the power-law photon index ($\Gamma$) -- at the average value of the parameters obtained from the best fitting results of the higher count rate spectra, $\Gamma$ = 1.6 and $N_{\rm H}$ = 3 $\times$ 10$^{21}$ cm$^{-2}$, and we only allowed the normalisation to be a free parameter in the fitting. The fitting results of all the individual \swift spectra are reported in Table~A1 and these were then used as the basis for further spectral analysis.

\begin{figure*}
\begin{center}

\includegraphics[width=15cm]{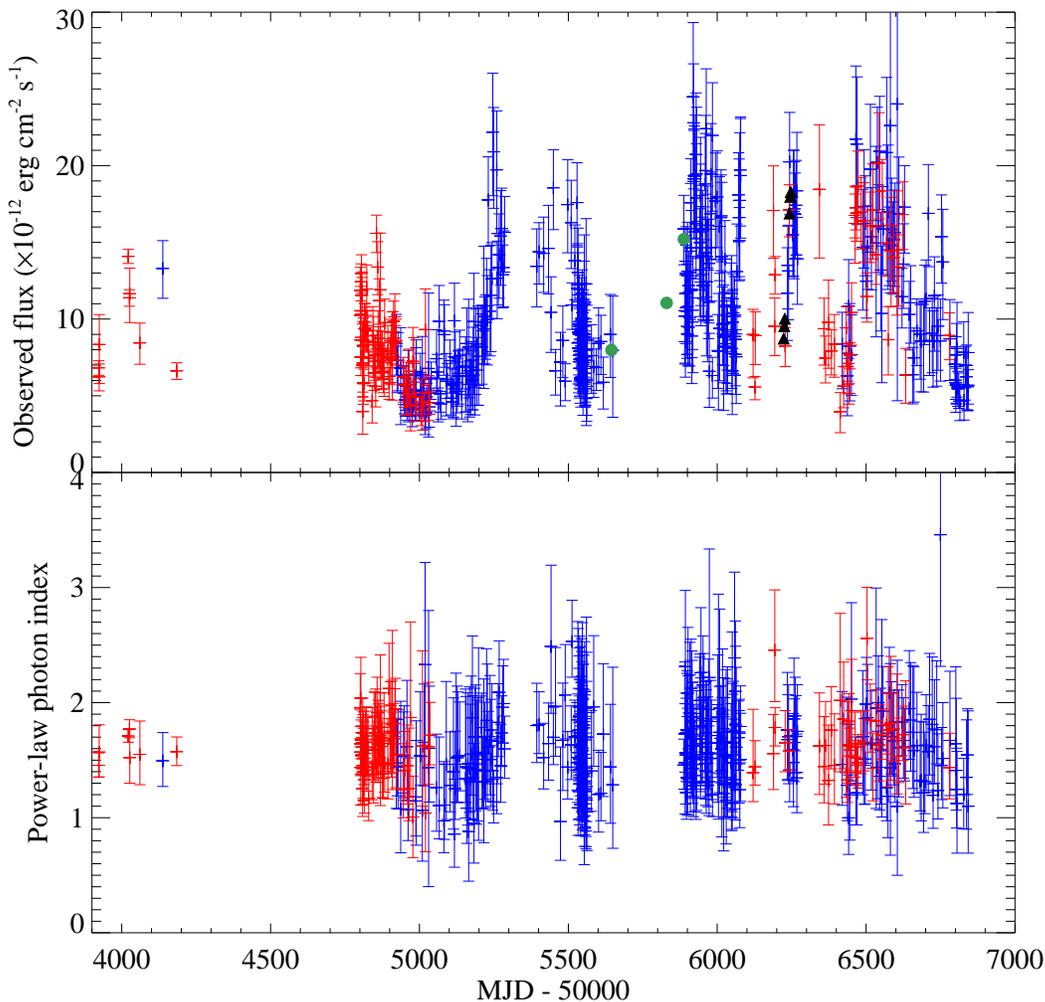}

\caption[]{{\it Top panel\/}: the light curve of Holmberg IX X-1, obtained from \swift on-axis (red) and off-axis (blue) observations. The time axis runs from June 2006 to August 2014.  The solid green circles indicate the 2011 \xmm observations, whilst the solid black triangles indicate the \xmm observations taken contemporaneously with \textit{NuSTAR} in 2012. {\it Bottom panel\/}: the photon index, plotted as a function of observing time, obtained by modelling the individual \swift spectra with an absorbed power-law model.}
\label{fig:swift_lc}
\end{center}
\end{figure*}

\begin{figure*}
\begin{center}

\includegraphics[width=15cm]{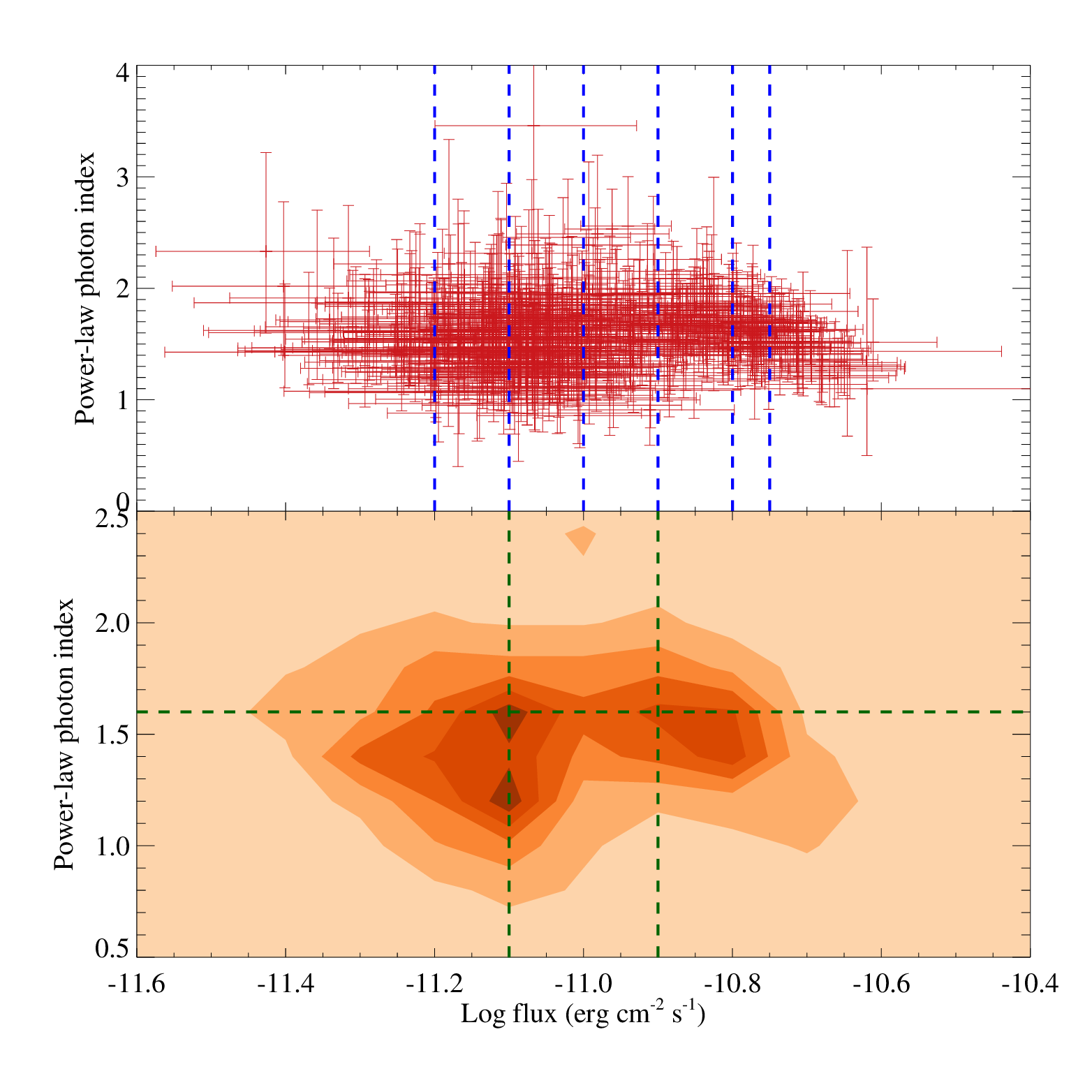}

\caption[]{{\it Top panel\/}: the hardness-intensity diagram, obtained by fitting the individual \swift spectra with an absorbed power-law model.  The boundaries used to segregate the \swift spectra into luminosity bins are plotted as blue dashed lines (see Section~\ref{sec:lum_bin_analysis}). {\it Bottom panel\/}: the contour plot of the data density corresponding to the plot above. The boundaries used to segregate the \swift spectra into the spectral index-luminosity bins are plotted as green dashed lines (see Section~\ref{sec:lum/index_bin_analysis}).}
\label{fig:hardness-intensity}
\end{center}
\end{figure*}

The light curve of \hoix obtained from the \swift monitoring observations is plotted in Fig.~\ref{fig:swift_lc}. It can be seen that the source flux varied by a factor of 4 -- 5 over a period of eight years.  As a first order analysis of the spectral evolution with time, we plot the photon index of the individual \swift spectra as a function of the observing time (bottom panel of Fig.~\ref{fig:swift_lc}); no obvious correlation between the photon index and observation time is evident.  To further examine the data, we plot the hardness-intensity diagram in Fig.~\ref{fig:hardness-intensity} (where we use the observed 0.3-10 keV flux as a proxy for the intensity, and the photon index for the hardness).  Interestingly, the spectra seem to have a bimodal distribution, with one peak at a flux of $\sim 8 \times 10^{-12} \rm ~erg~cm^{-2}~s^{-1}$ and the other peak at double this flux ($\sim 1.6 \times 10^{-11} \rm ~erg~cm^{-2}~s^{-1}$). We will consider this further in next section.

\subsubsection{Stacked spectral analysis: luminosity segregation} \label{sec:lum_bin_analysis}

It has been shown in previous studies of ULX spectra (e.g. \citealt{stobbart2006,gladstone2009}) that their spectral features, particularly the high energy curvature, are subtle and can only be seen in high quality spectra.  So, in order to obtain high S/N \swift spectra, we grouped the spectra into appropriate bins and stacked the spectra together to create a single spectrum for each bin, using the following criteria.

We began binning the individual \textit{Swift} spectra using luminosity criteria.  In particular, based on the bimodal distribution shown in Fig.~\ref{fig:hardness-intensity}, we divided the spectra in two main groups demarcated by an observed 0.3-10 keV flux of log $f_{\rm X} = -10.9$ (in units of erg cm$^{-2}$ s$^{-1}$), which we refer to hereafter as the low and high luminosity bins.  Given the large number of counts available we were able to further sub-divide the low luminosity bin into four bins: low1 ( log$f_{\rm X}$ $<$ -11.2), low2 (-11.2 $<$ log$f_{\rm X}$ $<$ -11.1), low3 (-11.1 $<$ log$f_{\rm X}$ $<$ -11) and low4 (-11 $<$ log$f_{\rm X}$ $<$ -10.9).  Similarly, the high luminosity bin was sub-divided into three bins: high1 (-10.9 $<$ log$f_{\rm X}$ $<$ -10.8), high2 (-10.8 $<$ log$f_{\rm X}$ $<$ -10.75) and high3 (log$f_{\rm X}$ $>$ -10.75); see Fig.~\ref{fig:hardness-intensity} (top panel) for the binning boundaries.  We stacked the spectra in each bin together using the script {\sc addspec}; the appropriate RMFs and ARFs were also combined by the script.

However, as the instrument response of the \swift XRT varies across the detector, combining responses with distinct differences  -- i.e. those of on-axis and off-axis spectra -- may risk introducing uncertainties into a combined response file for the stacked spectra.  To avoid this issue, we stacked the on- and off-axis spectra in each luminosity bin separately, such that we produced two spectra per bin.  Finally, we grouped the stacked spectra to have a minimum of 20 counts per bin within the spectra.  We note that we excluded two \swift observations -- observation IDs 00035335004 and 00035335005 -- from the following analysis since they have a high number of counts ($\ga$ 2900 counts) and so would potentially dominate the stacked spectrum they are each associated with (although they also have too few counts to be analysed separately). The properties of the stacked spectra in each luminosity bin are summarised in Table~\ref{tab:swift_spectra}.

    \begin{table}
      \centering
      \caption{The properties of the stacked \swift spectra}\label{tab:swift_spectra}
      \smallskip
      \begin{threeparttable}
          \begin{tabular}{lcccc}
\hline
Spectral bin$^{a}$ & \multicolumn{2}{c}{No. of spectra$^{b}$} & \multicolumn{2}{c}{Total counts$^{c}$}\\
			& on-axis & off-axis & on-axis & off-axis \\
\hline          
 \multicolumn{5}{c}{Luminosity binning} \\
Low1(Low)		&	23	&	60	&	4494	&	6374	\\
Low2(Low)		&	29	&	58	&	10069	&	7764	\\
Low3(Low)		&	36	&	89	&	11275	&	14318	\\
&&&&\\
Low4(Medium)		&	13	&	45	&	5804	&	8625	\\
&&&&\\								
High1(High)		&	10	&	51	&	5372	&	13287	\\
High2(High)		&	11	&	31	&	6587	&	8661	\\
High3(High)		&	4	&	33	&	1574	&	9564	\\

\hline	
 \multicolumn{5}{c}{Photon index and luminosity binning} \\	

 \multicolumn{5}{c}{{\it Hard spectral bin}} \\							
Low		&	23	&	46	&	8202	&	6158	\\
Medium	&	20	&	78	&	6372	&	12152	\\
High	&	11	&	56	&	7704	&	15736	\\

\multicolumn{5}{c}{{\it Soft spectral bin}} \\							
Low	&	23	&	31	&	5898	&	4549	\\
Medium	&	26	&	55	&	10495	&	10704	\\
High	&	14	&	55	&	5829	&	15480	\\             
\hline             
             
         \end{tabular}
         \begin{tablenotes}
         \item \textbf{Note.} $^{a}$The names of the stacked \swift spectral bins. Those in brackets indicate the new names after the spectra are re-binned to further improve S/N (see Section~\ref{sec:lum_bin_analysis}).  $^{b}$The number of \swift spectra contributing to each spectral bin.  $^{c}$The total number of photon counts from all spectra contributing to the spectral bin.
         \end{tablenotes}
      \end{threeparttable}
    \end{table}

We began the analysis of the stacked spectra by modelling them with the absorbed MCD plus Comptonisation model. We modelled the on- and off-axis spectra of each luminosity bin simultaneously, adding a multiplicative constant to the model to allow for any calibration differences between the two spectra.  The parameter was frozen at unity for the on-axis spectra, whilst that of the off-axis spectra was allowed to be a free parameter; we found that the disagreement between the spectra was at no more than the $\la$ 10\% level.

\begin{figure}
\begin{center}

\includegraphics[width=8.5cm]{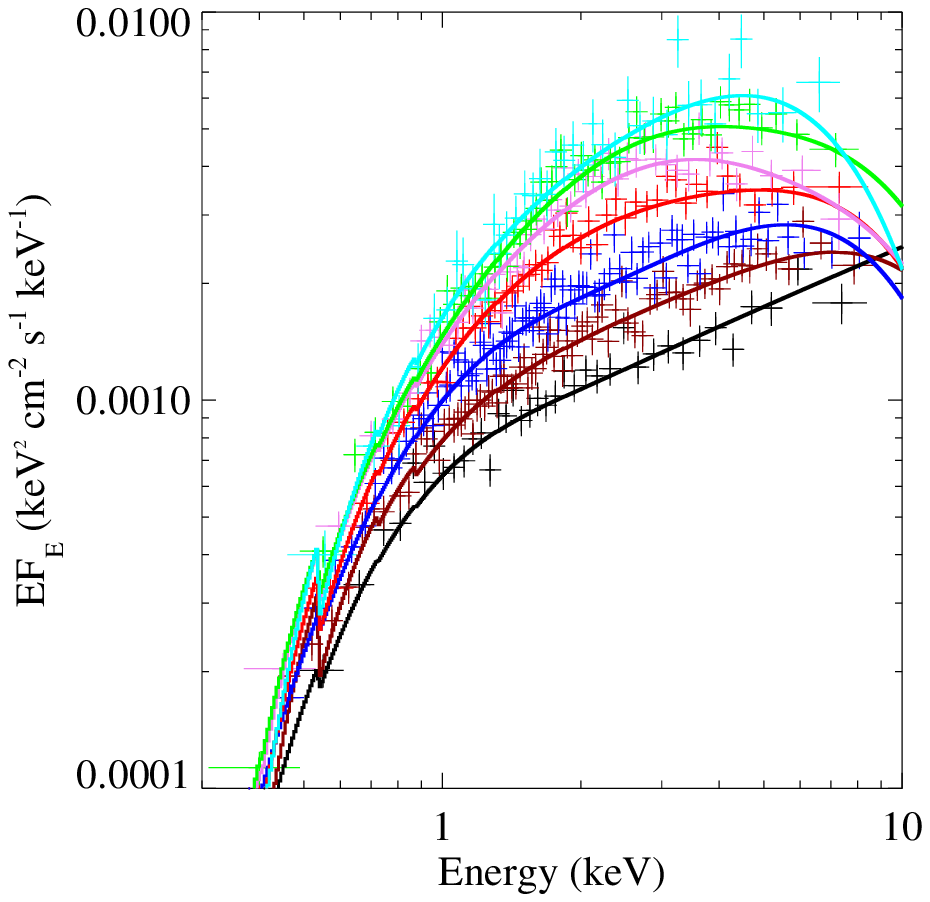}

\caption[]{The stacked \swift spectra, showing (from bottom to top) the low1 (black), low2 (brown), low3 (blue), low4 (red), high1 (violet), high2 (green) and high3 (cyan) luminosity bins.  Only on-axis spectra are shown, and they are re-binned to a minimum of 10$\sigma$ statistical significance (5$\sigma$ statistical significance for the high3 luminosity bin spectrum) per data point to clarify the plot.  The best-fitting absorbed MCD plus \comptt model for each spectrum is also shown by the  solid lines.}
\label{fig:swift_spectra_7bin_spectra}
\end{center}
\end{figure}

The stacked spectra along with the best-fitting models are shown in Fig.~\ref{fig:swift_spectra_7bin_spectra}. The plot reveals an interesting result; it is obvious that the spectra change shape as the luminosity increases -- especially in the $\sim$1-6 keV energy band -- from appearing as flat spectra at low luminosity to becoming more curved as the luminosity increases.  This demonstrates clear evolution of the average ULX spectra with increasing luminosity.  However, we found that the data were not sufficiently high quality to strongly constrain the model parameters with this choice of binning.  So, in order to improve the constraints on the spectral fitting, we increased the S/N of the stacked spectra by combining those with similar characteristics in the $\sim$1-6 keV energy band;  hence we stacked the spectra of the low1, low2 and low3 luminosity bins together as these spectra are similarly flat (hereafter the low luminosity bin; see Table~\ref{tab:swift_spectra}). In contrast, the spectra of high1, high2 and high3 luminosity bins seem to be very curved and so were stacked together (hereafter the high luminosity bin). The spectra of the low4 luminosity bin was left to represent an intermediate spectral stage between the flat and curved spectra; hereafter we refer to it as the medium luminosity bin.


    \begin{table*}

      \caption{The best fitting parameters for the stacked \swift spectra modelled by a MCD plus Comptonisation model}\label{tab:swift_fitting_result}
      \smallskip
      \begin{threeparttable}
          \begin{tabular}{lccccccc}
          \hline
             Spectral bin & $N_{\rm H}$$^{a}$ & $kT_{\rm in}$$^{b}$ & $kT_{\rm e}$$^{c}$& $\tau_{\rm e}$$^{d}$& $\chi^{2}$/d.o.f.$^{e}$ & $L_{\rm X}$$^{f}$ & $f_{\rm MCD}/f_{\rm tot}$$^{g}$\\
             & & (keV) &(keV)  & & & & \\
             \hline
 \multicolumn{7}{c}{Luminosity binning} \\             
Low 	& $	0.19	_{	-0.01	} ^{ +	0.03	} $ & $	<0.11					 $ & $	2.61	_{	-0.28	} ^{ +	0.41	} $ & $	7.45	_{	-0.59	} ^{ +	0.52	} $ & $	1028.85	/	970	$&	$	10.42	_{-	0.16	}^{+	0.14	}$ & $	0.00	$$^{*}$ 	\\
Medium	& $	0.12	_{	-0.07	} ^{ +	0.02	} $ & $	0.81	_{-0.51} ^{+0.41}				$& $	1.89	_{	-0.51	} ^{ +	1.04	} $ & $		>	7.15			$ & $	441.40	/	482	$&	$	15.43	_{-	0.64	}^{+	0.67	}$ & $	<0.60	$ 	\\
High	& $	0.12	\pm0.01 $ & $	0.73	_{	-0.15	} ^{ +	0.96	} $ & $	1.87	_{	-0.30	} ^{ +	81.77	} $ & $		>	1.57			 $ & $	819.04	/	849	$&	$	20.19	_{-	0.22	}^{+	0.48	}$ & $	0.37	_{-	0.07	}^{+	0.12	}$ 	\\

\hline

 \multicolumn{7}{c}{Photon index and luminosity binning} \\   
  \multicolumn{7}{c}{{\it Hard spectral bin}} \\   
Low	& $	0.21	_{	-0.05	} ^{ +	0.09	} $ & $	0.16	_{	-0.02	} ^{ +	0.04	} $ & $	2.44	_{	-0.39	} ^{ +	0.84	} $ & $	8.45	_{	-1.41	} ^{ +	1.24	} $ & $	496.78	/	501	$&	$	9.19	_{-	0.32	}^{+	0.34	}$ & $	0.13	_{-	0.11	}^{+	0.05	}$ 	\\
Medium	& $	0.20 \pm0.09$ & $	0.14	_{	-0.03	} ^{ +	0.04	} $ & $	1.96	_{	-0.20	} ^{ +	0.29	} $ & $	9.53	_{	-0.98	} ^{ +	1.02	} $ & $	614.57	/	580	$&	$	12.61	_{-	0.40	}^{+	0.42	}$ & $	<0.20					$ 	\\
High	& $	0.11	_{	-0.01	} ^{ +	0.02	} $ & $	0.82	_{	-0.28	} ^{ +	0.76	} $ & $		>	1.03			$ & $		>	0.97			 $ & $	706.40	/	665	$&	$	20.47	_{-	0.66	}^{+	0.77	}$ & $	0.40	_{-	0.26	}^{+	0.32	}$ 	\\

\multicolumn{7}{c}{{\it Soft spectral bin}} \\   

Low	& $	0.10	\pm0.02$ & $	0.85	_{	-0.39	} ^{ +	0.12	} $ & $		>	1.89			$ & $		>	3.43			$ & $	312.87	/	391	$&	$	9.78	_{-	0.50	}^{+	0.62	}$ & $	0.53	_{-	0.21	}^{+	0.04	}$ 	\\
Medium	& $	0.13	\pm0.02	 $ & $	0.78	_{	-0.34	} ^{ +	0.31	} $ & $	2.24	_{	-0.59	} ^{ +	3.46	} $ & $	9.56	_{	-3.06	} ^{ +	67.50	} $ & $	621.86	/	601	$&	$	13.20	_{-	0.44	}^{+	0.51	}$ & $	0.50	_{-	0.23	}^{+	0.09	}$ 	\\
High	& $	0.13	_{	-0.03	} ^{ +	0.02	} $ & $	0.68	_{	-0.19	} ^{ +	0.37	} $ & $		>	1.23			$ & $		>	0.17			$ & $	567.48	/	588	$&	$	19.88	_{-	0.71	}^{+	0.78	}$ & $	0.34	_{-	0.11	}^{+	0.10	}$ 	\\

              \hline
         \end{tabular}
         \begin{tablenotes}
         \item \textit{Note.} $^{a}$Absorption column external to our Galaxy, in units of 10$^{22}$ cm$^{-2}$. $^{b}$The inner disc temperature of the MCD component. $^{c}$The plasma temperature of the Comptonising corona. $^{d}$The optical depth of the Comptonising corona. $^{e}$Minimum $\chi^{2}$ over degrees of freedom. $^{f}$Observed X-ray luminosity in the 0.3-10 keV energy band, in units of 10$^{39}$ erg s$^{-1}$. $^{g}$The ratio of MCD flux to total unabsorbed X-ray flux in the 0.3-10 keV energy band. $^{*}$No flux contribution from the MCD component in the best-fitting model to this dataset.
          
                  \end{tablenotes}
      \end{threeparttable}
    \end{table*}

\begin{figure}
\begin{center}

\includegraphics[width=8.5cm]{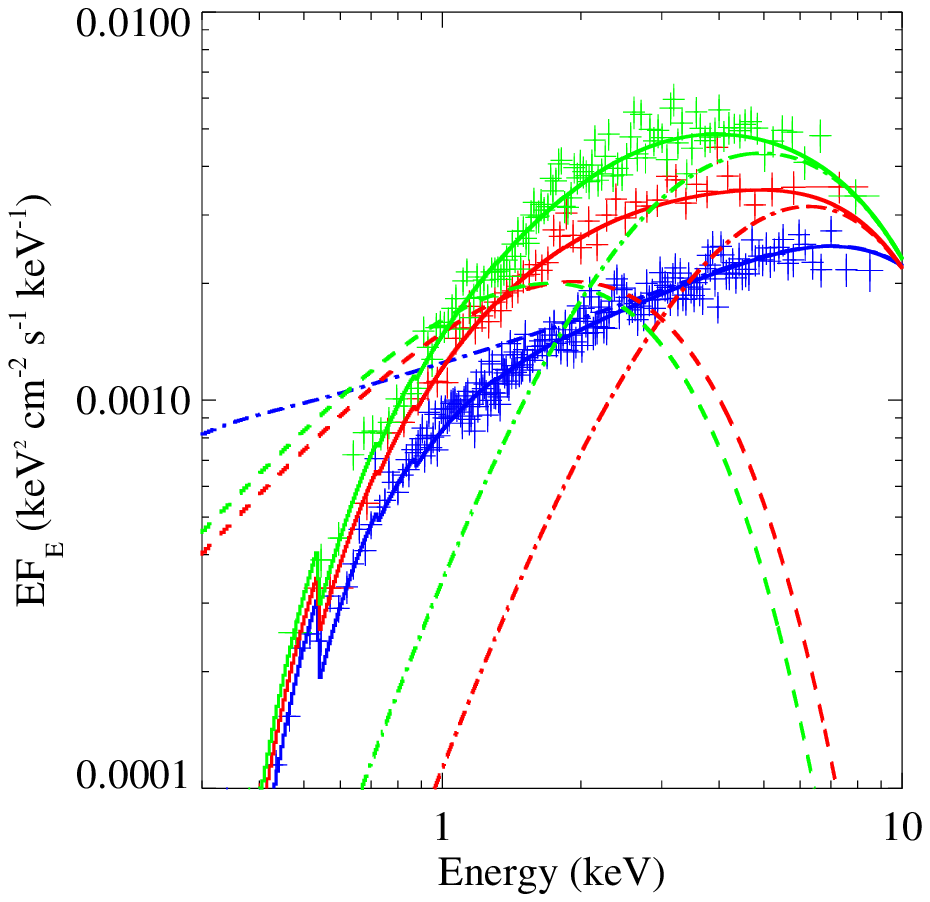}

\caption[]{The stacked \swift spectra in (from bottom to top) the low (blue), medium (red) and high (green) luminosity bins. Only on-axis spectra are shown and they are re-binned to a minimum of 10$\sigma$ statistical significance for clarity.  The best-fitting absorbed MCD plus \comptt models are shown as solid lines; the individual components, corrected for absorption, are also plotted as the dashed (MCD) and dash-dotted lines (Comptonisation).}
\label{fig:swift_lum_bin_spectra}
\end{center}
\end{figure}

Again, we analysed the stacked spectra in the low, medium and high luminosity bins by modelling them with a MCD plus Comptonisation model; this produced statistically acceptable fits in all cases (null hypothesis probability $> 0.05$), and the best fitting results are reported in Table~\ref{tab:swift_fitting_result} and shown in Fig~\ref{fig:swift_lum_bin_spectra}. Overall, similar spectral evolution to the previous, finer flux segregation is found; the average spectra are flat at low luminosity and become more curved at higher luminosity.  They appear most curved at the highest ULX luminosity.  Hence it appears that the two peaks in Fig.~\ref{fig:hardness-intensity} occur where the spectra appear different; flat in the low luminosity bin, and curved in the medium and high luminosity bins.

We can further examine the changes in the average spectra by looking at what happens to the two components in the spectral fits.  At low luminosity, the spectral fit is dominated by the \comptt component; the MCD component makes a no contribution, resulting in the flat spectrum.  As the luminosity increases, the MCD component emerges to dominate the soft end of the medium and high luminosity spectra, whilst the \comptt component dominates the hard end.  However, this MCD appears very consistent between the medium and high bins, whereas the \comptt appears to continue to change, with its peak evolving towards a lower temperature as the luminosity increases in all 3 spectra (although we note that the parameter values for the fits to the three datasets shown in Table~\ref{tab:swift_fitting_result} are formally indistinguishable at the 90\% level, with the exception of the higher absorption column in the low luminosity spectrum, and the MCD changing from absent to providing between roughly a third and a half of the flux in the high luminosity spectrum).  It is this apparent cooling of the corona, combined with the emergence of the MCD, that leads to the increasingly curved spectra at higher luminosities.

\subsubsection{Stacked spectral analysis: photon index and luminosity segregation}
\label{sec:lum/index_bin_analysis}

The hardness-intensity diagram in Fig.~\ref{fig:hardness-intensity} shows that the distribution of the data has definite structure. In the previous section, we examined the spectral evolution based on luminosity criteria alone. However, the hardness-intensity diagram also demonstrates that the photon indexes of individual \swift spectra vary widely for the same luminosity, between $\sim1 - 2.5$.  Indeed, there is even some evidence in the bottom panel of Fig.~\ref{fig:hardness-intensity} for the low luminosity density peak having two distinct sub-peaks, either side of $\sim \Gamma = 1.4$.  This strongly suggests that the average spectra are not a simple function of luminosity.  Thus, in this section, we re-examine the \swift spectra using new binning criteria that combine the observed luminosities and spectral indexes of the individual observations. Firstly, we divided the spectra into hard ($\Gamma$ $<$ 1.6) and soft   ($\Gamma$ $>$ 1.6) spectral bins.  Then we further segregate the spectra in these bins into three luminosity bins: low (log$f_{\rm X}$ $<$ -11.1), medium (-11.1 $<$ log$f_{\rm X}$ $<$ -10.9) and high (log$f_{\rm X}$ $>$ -10.9), whose boundaries are shown in the bottom panel of Fig.~\ref{fig:hardness-intensity}.  The criteria result in six spectral bins, the properties of which are summarised in Table~\ref{tab:swift_spectra}. 

\begin{figure*}
\begin{center}

\includegraphics[width=8.5cm]{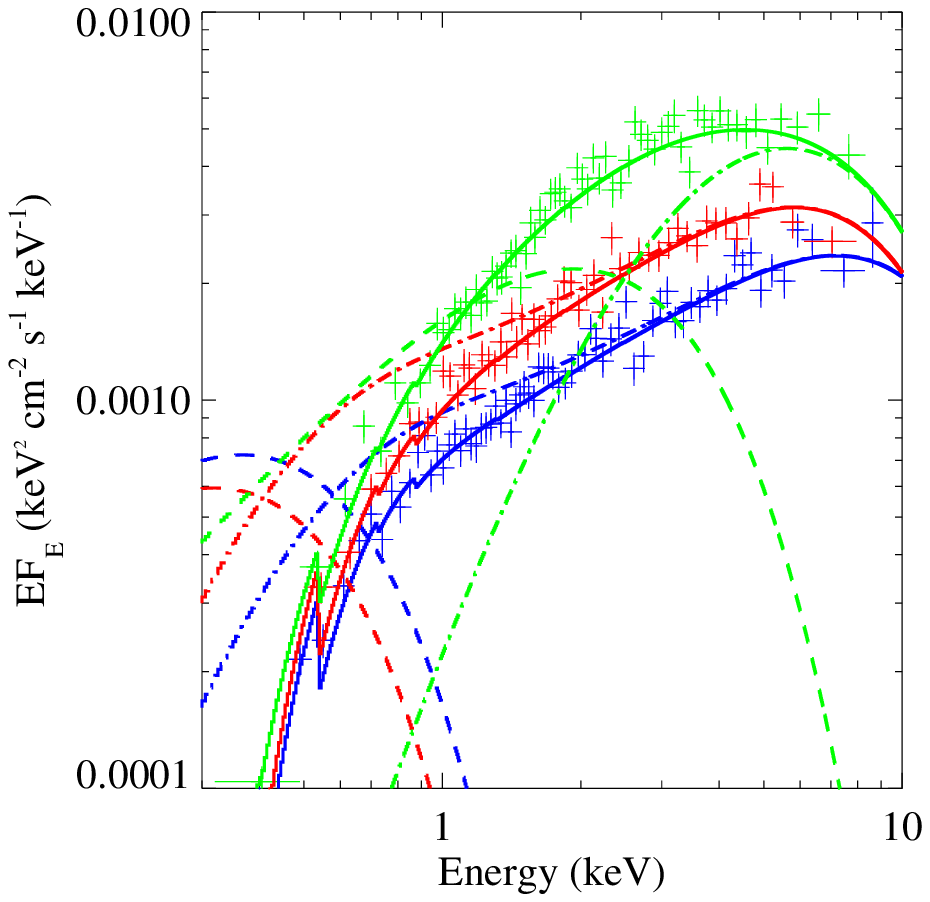}~\includegraphics[width=8.5cm]{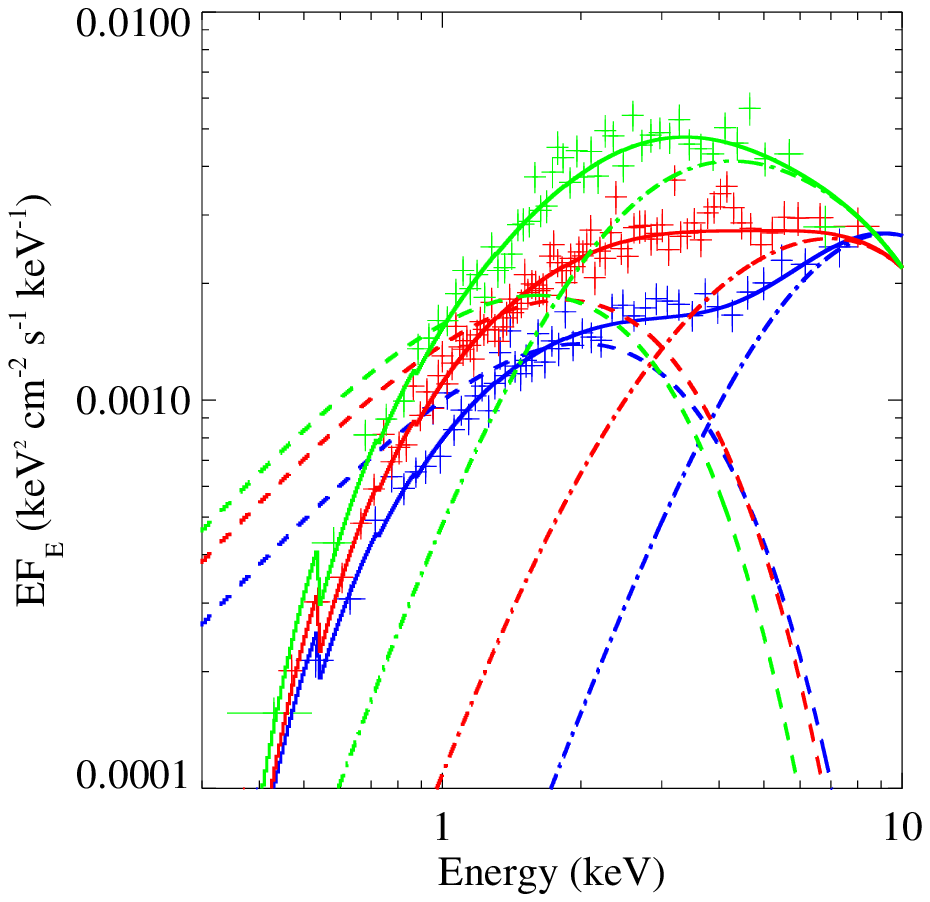}

\caption[]{The stacked \swift spectra binned by a combination of spectral index and luminosity criteria. {\it Left panel\/}: the stacked hard spectra. {\it Right panel\/}: the stacked soft spectra.  Only on-axis spectra are shown and they are re-binned to a minimum of 10$\sigma$ statistical significance per data point to clarify the plot.  The line styles, colours and order of the spectra are as per Fig.\ref{fig:swift_lum_bin_spectra}.} 
\label{fig:swift_hardness-lum_bin_spectra}
\end{center}
\end{figure*}

We stacked the spectra in each bin together using the script {\sc addspec}.  Again, for each spectral bin, we stacked the on- and off-axis spectra separately. We note that, in this section, we do not include the low count \swift spectra in the analysis (the 55 observations that have no more than between 50 and 100 counts); as these spectra have an assumed spectral index of 1.6 (see Section~\ref{sec:swift individual spectra}), they cannot be assimilated into either the hard or soft spectral bins.  Finally, we grouped the stacked spectra to have a minimum of 20 counts per bin within each spectrum, and analysed the evolution of the spectra using the same model as in Section~\ref{sec:lum_bin_analysis}.  The best fitting results are shown in the lower panel of Table~\ref{tab:swift_fitting_result}.   We illustrate the spectral evolution with increasing luminosity, for the soft and hard bins separately, in Fig~\ref{fig:swift_hardness-lum_bin_spectra}. 

The results obtained from this binning method are very interesting; the spectra in the hard and soft spectral bins show different evolution with increasing luminosity.  For the hard spectra, the overall spectral evolution in shape is broadly consistent with that of the pure luminosity binning shown Section~\ref{sec:lum_bin_analysis}; the spectra seem to be flat in the $\sim 1 - 6$ keV range at low luminosity, and to become curved at high luminosity.  The detailed evolution of the two components is, however, somewhat different.  The MCD component contributes at roughly the same, low level to the total spectra at low and medium luminosity ($< 20\%$ of the total unabsorbed flux).  However, at high luminosity, the component becomes very significantly hotter (cf. Table~\ref{tab:swift_fitting_result}) and dominates the soft end of the spectrum, contributing $\sim 40\%$ of the total flux (although again we cannot formally exclude the lower limit on this contribution from overlapping with the low and medium bin contributions).  In contrast, the \comptt component is very dominant at low and medium luminosity; however at high luminosity, the component only dominates the hard band.  We do, however, see the same apparent gradual decrease in temperature in the hard component as the total luminosity increases as for the pure luminosity binning, but again have the same caveat about a lack of formal distinction at the 90\% significance level. 

In contrast, the evolution of the soft spectra with luminosity is somewhat different from that of the hard spectra, in particular at low and medium luminosity (see right panel of Fig.~\ref{fig:swift_hardness-lum_bin_spectra}).  Overall, the spectrum appears to be distinctly two component in the low luminosity bin; however, as the luminosity increases, the individual components of the MCD and \comptt seem to merge together, resulting in a single-component-like curved spectrum.  Furthermore, unlike the hard spectra, the MCD component always appears to dominate the softer energy band of the spectra, contributing $\sim 35 - 50\%$ of the total unabsorbed flux, and remains roughly constant in temperature as the luminosity increases (i.e. does not evolve very dramatically with increasing total luminosity).  These MCD components are distinctly different from those of the hard spectra, showing significantly higher fractional contributions to the overall flux, and temperatures, than the MCD components in the corresponding hard spectra.  In the soft spectra the \comptt component also shows possible temperature evolution, displaying the same pattern in apparent decreasing peak energy as the luminosity increases as is seen in the other analyses, although this again remains unconfirmed within the uncertainty limits on the parameter fitting.  

\subsection{\xmm spectral analysis}
\label{sec:xmm spectral analysis}

Unlike the \swift spectra, the 13 \xmm spectra were of sufficiently high quality to enable us to study \hoix without stacking.  We analysed these spectra using the same model as for the \swift stacked spectra in Section~\ref{sec:lum_bin_analysis} and \ref{sec:lum/index_bin_analysis}.  For each individual observation, we model the PN, MOS1 and MOS2 spectra simultaneously, adding a multiplicative constant to the model to correct for any residual calibration differences between the \xmm detectors; in practise this offset is at the $\la 10$ per cent level. The best fitting results are tabulated in Table~\ref{tab:xmm_fitting_result}; to facilitate comparison with the other analyses, we sort the spectra in ascending order of their observed luminosities and also classify each \xmm spectrum into an appropriate luminosity group, corresponding to the boundaries defined in Section~\ref{sec:lum_bin_analysis}.  Acceptable fits were found in all but 2 cases; observation ID 0200980101 is very marginally rejected (null hypothesis probability of 0.04), and 0111800101 was rejected with a null hypothesis of $\sim 5 \times 10^{-3}$, i.e. a $\sim 3\sigma$ confidence rejection.  However, it is notable that these observations are the highest quality data for \hoix and that there is a suggestion of residuals in the soft X-ray emission, in common with other ULXs, that may be causing this result \citep{middleton2015b}.


    \begin{table*}
      \centering
      \caption{The best fitting results for the \xmm spectra modelled by a MCD plus Comptonisation model}\label{tab:xmm_fitting_result}
      \smallskip
      \begin{threeparttable}
          \begin{tabular}{lcccccccc}
          \hline
             Obs. ID & $N_{\rm H}$$^{a}$ & $kT_{\rm in}$$^{b}$ & $kT_{\rm e}$$^{c}$ & $\tau_{\rm e}$$^{d}$& $\chi^{2}$/d.o.f.$^{e}$  & $L_{\rm X}$$^{f}$ & $f_{\rm MCD}/f_{\rm tot}$$^{g}$ & Group$^{h}$\\
            & & (keV) &(keV)  & & & & & \\
             \hline
0200980101	& $	0.11	\pm0.01	$ & $	0.26	\pm0.02	$ & $	2.25	_{	-0.13	} ^{ +	0.15	} $ & $	9.63	_{	-0.54	} ^{ +	0.61	} $ & $	(536.48	/	482)	$&	$	8.50	\pm	0.09			$ & $	0.19	\pm	0.01	$	&	Low 	\\
0112521001	& $	0.11	\pm0.02$ & $	0.25	\pm0.05 $ & $	2.57	_{	-0.45	} ^{ +	0.94	} $ & $	7.54	_{	-1.40	} ^{ +	1.39	} $ & $	324.22	/	336	$&	$	9.83	\pm	0.24			$ & $	0.16	_{-	0.05	}^{+	0.03	}$	&	Low 	\\
0657802001	& $	0.16	_{	-0.04	} ^{ +	0.05	} $ & $	0.24	_{	-0.04	} ^{ +	0.05	} $ & $		>	2.41			 $ & $	0.75	_{	-0.61	} ^{ +	38.69	} $ & $	200.62	/	205	$&	$	11.17	_{-	0.47	}^{+	0.48}	$ & $	0.19	_{-	0.04	}^{+	0.03	}$	&	Low	\\
0112521101	& $	0.10	_{	-0.06	} ^{ +	0.03	} $ & $	0.22	_{	-0.04	} ^{ +	0.06	} $ & $	2.64	_{	-0.42	} ^{ +	0.72	} $ & $	6.95	_{	-0.95	} ^{ +	1.04	} $ & $	368.12	/	350	$&	$	11.27	\pm	0.23			$ & $					<0.17	$	&	Low 	\\
0693850801	& $	0.10	_{	-0.06	} ^{ +	0.03	} $ & $	0.22	\pm0.04	 $ & $		>	3.15			 $ & $	1.07	_{	-1.02	} ^{ +	4.88	} $ & $	382.71	/	347	$&	$	12.24	\pm	0.25			$ & $	<0.11$	&	Low 	\\
0693850901	& $	0.11	_{	-0.07	} ^{ +	0.03	} $ & $	0.26	_{	-0.07	} ^{ +	0.10	} $ & $	2.46	_{	-0.45	} ^{ +	0.67	} $ & $	7.61	_{	-1.26	} ^{ +	1.86	} $ & $	362.56	/	364	$&	$	13.38	\pm	0.31			$ & $	<0.18$	&	Low 	\\
0693851001	& $	0.10	_{	-0.06	} ^{ +	0.02	} $ & $	0.30	_{	-0.08	} ^{ +	0.20	} $ & $	2.16	_{	-0.46	} ^{ +	0.52	} $ & $	8.27	_{	-1.35	} ^{ +	4.61	} $ & $	328.58	/	329	$&	$	14.05	_{-	0.33	}^{+	0.34}	$ & $	0.17	_{-	0.16	}^{+	0.07	}$	&	Medium	\\
0657801801	& $	0.08	_{	-0.07	} ^{ +	0.03	} $ & $	0.28	_{	-0.04	} ^{ +	0.11	} $ & $	2.56	_{	-0.43	} ^{ +	0.98	} $ & $	6.89	_{	-1.23	} ^{ +	1.17	} $ & $	331.12	/	354	$&	$	15.47	_{-	0.40	}^{+	0.41}	$ & $					<0.24	$	&	Medium	\\
0111800101	& $	0.13	\pm0.01	 $ & $	0.52	_{	-0.18	} ^{ +	0.33	} $ & $	1.30	_{	-0.11	} ^{ +	0.09	} $ & $		>	11.39			$ & $	(363.79	/	297)	$&	$	19.20	_{-	0.31	}^{+	0.32}	$ & $	0.36	_{-	0.13	}^{+	0.10	}$	&	High 	\\
0657802201	& $	0.10	_{	-0.04	} ^{ +	0.03	} $ & $	0.28	_{	-0.03	} ^{ +	0.07	} $ & $	1.97	_{	-0.13	} ^{ +	0.16	} $ & $	8.38	_{	-0.48	} ^{ +	0.61	} $ & $	391.12	/	415	$&	$	21.28	_{-	0.31	}^{+	0.32}	$ & $	0.09	_{-	0.06	}^{+	0.04	}$	&	High 	\\
0693851701	& $	0.11	_{	-0.02	} ^{ +	0.01	} $ & $	0.71	\pm0.17	 $ & $	1.76	_{	-0.13	} ^{ +	0.17	} $ & $	11.77	_{	-1.68	} ^{ +	3.83	} $ & $	432.58	/	406	$&	$	23.65	_{-	0.35	}^{+	0.34}	$ & $	<0.47	$	&High \\
0693851801	& $	0.07	_{	-0.06	} ^{ +	0.04	} $ & $	0.26	_{	-0.03	} ^{ +	0.07	} $ & $	1.54	_{	-0.05	} ^{ +	0.06	} $ & $	11.43	_{	-0.41	} ^{ +	0.47	} $ & $	428.36	/	407	$&	$	25.17	\pm	0.33			$ & $					<0.12	$	&	High \\
0693851101	& $	0.11	\pm0.02	 $ & $	0.77	_{	-0.24	} ^{ +	0.41	} $ & $	1.92	_{	-0.22	} ^{ +	0.38	} $ & $		>	8.57			 $ & $	384.06	/	355	$&	$	25.62	\pm	0.55			$ & $	0.44	_{-	0.11	}^{+	0.10	}$	&	High 	\\
             \hline
         \end{tabular}
         \begin{tablenotes}
         \item \textit{Note.}$^{a}$Absorption column external to our Galaxy, in units of 10$^{22}$ cm$^{-2}$. $^{b}$The inner disc temperature of the MCD component. $^{c}$The plasma temperature of the Comptonising corona. $^{d}$The optical depth of the Comptonising component. $^{e}$Minimum $\chi^{2}$ over degrees of freedom.  Parentheses highlight fits for which the null hypothesis probability is $< 0.05$.  $^{f}$Observed X-ray luminosity in the 0.3-10 keV energy band, in units of 10$^{39}$ erg s$^{-1}$. $^{g}$The ratio of MCD flux to the total unabsorbed X-ray flux in the 0.3-10 keV energy band. $^{h}$The luminosity bin of the \xmm spectra. We classify the spectra using the same luminosity boundaries used to create the \swift low, medium and high luminosity spectra in Section~\ref{sec:lum_bin_analysis}.
         \end{tablenotes}
      \end{threeparttable}
    \end{table*}


\begin{figure*}
\begin{center}

\includegraphics[width=6cm]{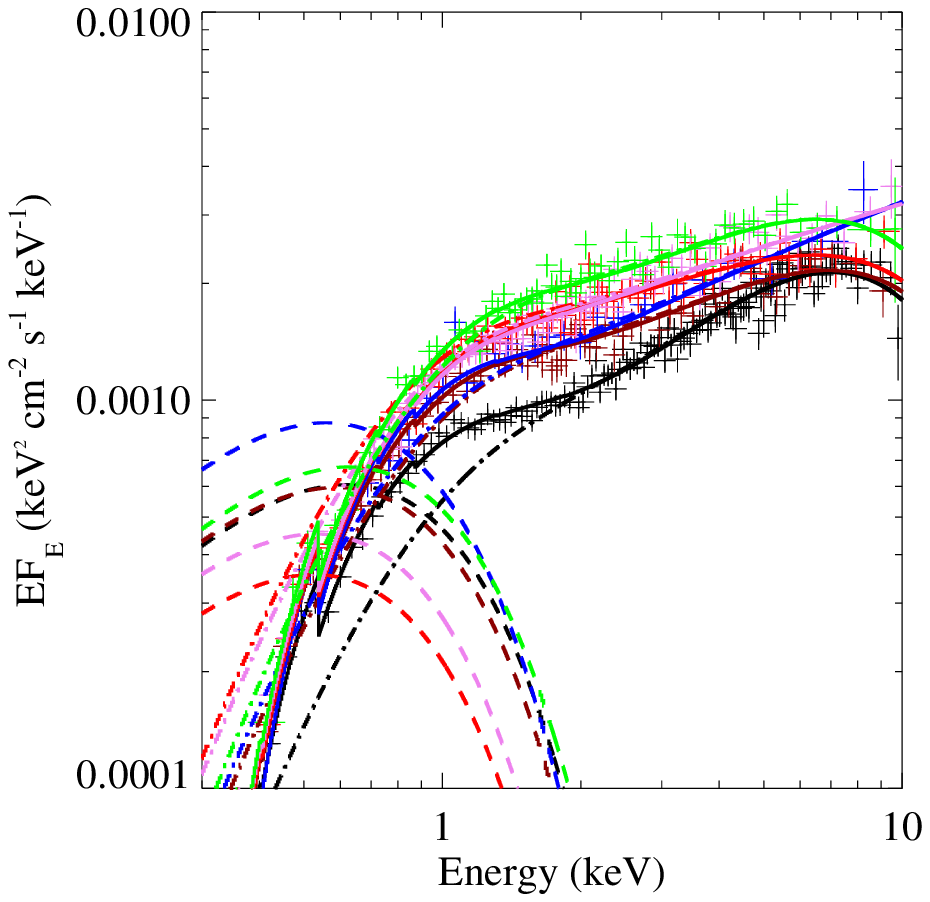}~\includegraphics[width=6cm]{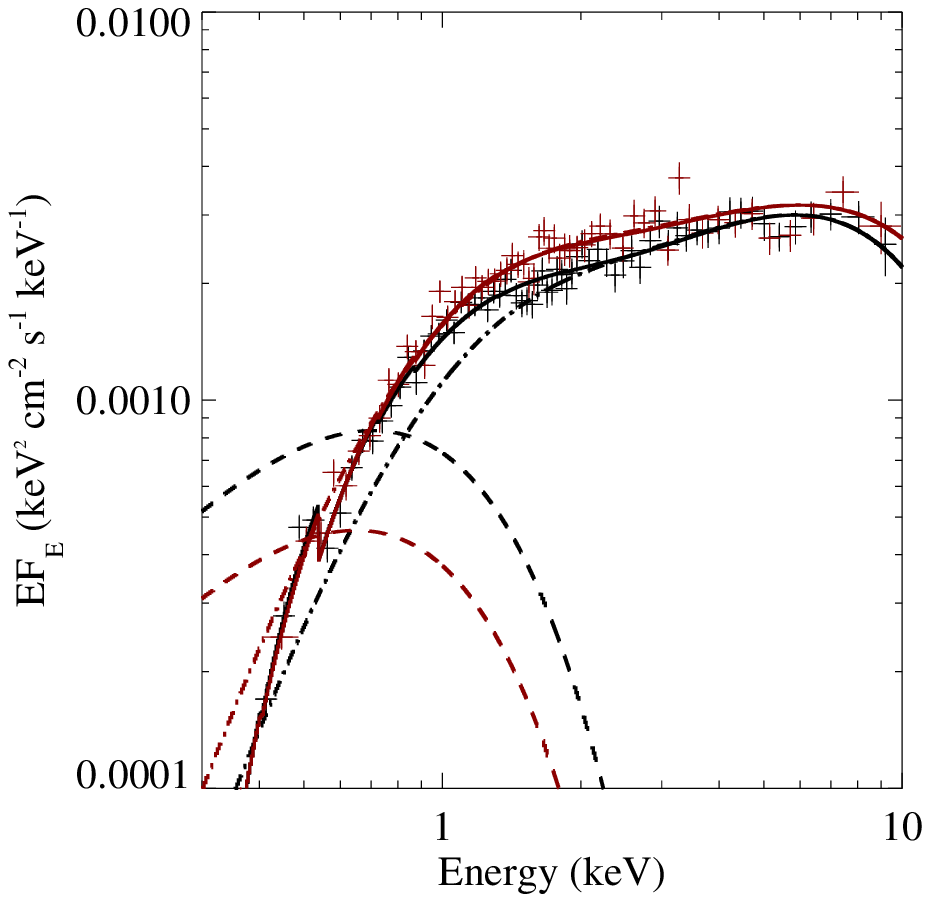}~\includegraphics[width=6cm]{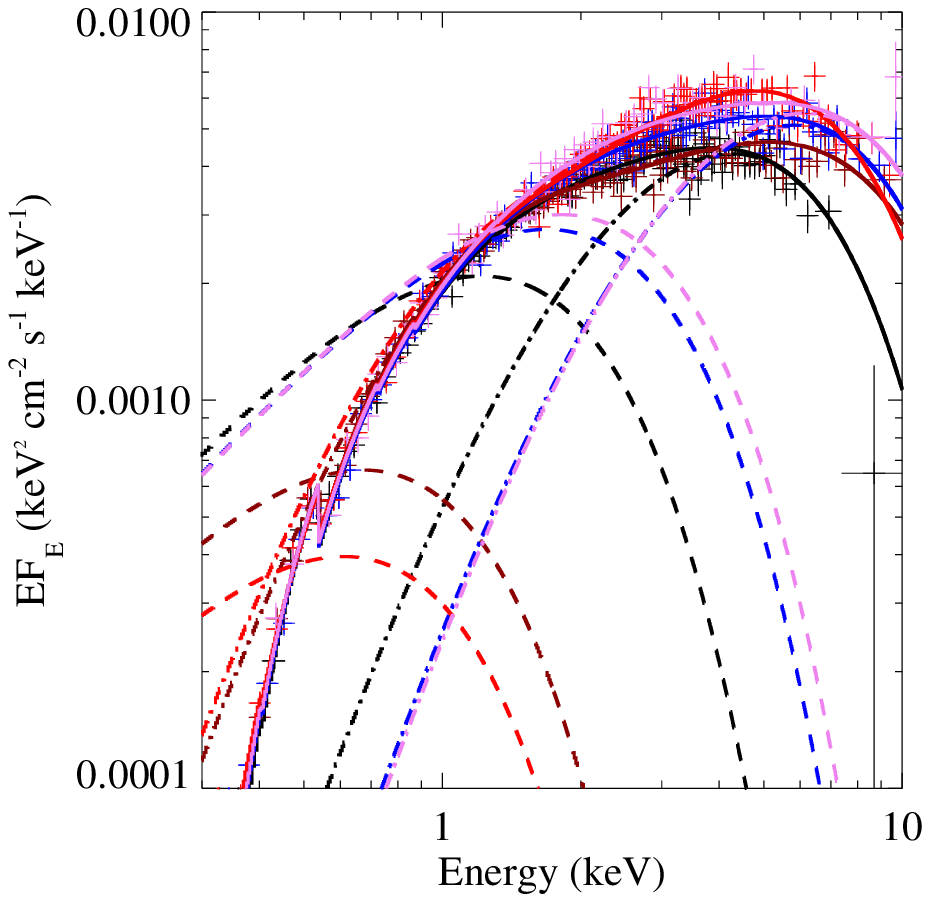}

\caption[]{The \xmm spectra, segregated into low ({\it left panel}), medium ({\it middle panel}) and high ({\it right panel\/}) luminosity groups using the same luminosity boundaries used to create the stacked \swift spectra.  Only PN spectra (or MOS1 in case of ObsID. 0111800101) are shown and they are re-binned to a minimum of 10$\sigma$ statistical significance for clarity.  The best-fitting absorbed MCD plus \comptt model is also shown as solid lines; the individual components of MCD and Comptonisation, corrected for absorption, are also plotted as dashed lines and dash-dotted lines, respectively. For each panel, the colour coding is used to indicate the relative luminosity of the source, ascending in order of increasing luminosity from black to brown, blue, red, violet and green.}
\label{fig:xmm_spectra}
\end{center}
\end{figure*}

We show the spectral evolution of \hoix with luminosity, as seen by {\it XMM-Newton\/}, in Fig.~\ref{fig:xmm_spectra}.  Overall, we see some broad consistencies with the \swift stacked spectra.  There are some obvious degeneracies in spectral shape with luminosity, as suggested by splitting the stacked \swift spectra into hard and soft bins, particularly in the low and high luminosity \xmm groups.  In the low luminosity group this manifests as both distinct two-component and flat spectral shapes (with the flat spectra apparently slightly more luminous); in the high luminosity group the differences are mainly in the turnover of the hard component.  However, we again see the same trends in the data, with mostly flatter data (in the 1-6 keV range) giving way to much more pronounced curvature as the luminosity of \hoix increases.

The evolution of the two individual components is also interesting.  In the low and medium luminosity groups, only a small fraction of the total spectrum originates in the MCD component (generally $< 20\%$), and its properties remain approximately constant (e.g. $kT_{\rm in}$ is invariant within errors; see Table~\ref{tab:xmm_fitting_result}) as the luminosity increases from low to medium luminosity.  However, in the high luminosity group, this component contributes significantly more to the spectrum in 3/5 cases ($\sim 40\%$ of the total unabsorbed flux), and in these cases is significantly warmer ($\sim 0.7$ keV as opposed to $\sim 0.25$ keV);  however there remain two cases -- observation ID 0657802201 and 0693851801 -- in which the MCD temperatures and contributions are as low as those of the low luminosity bin.  The \comptt dominates the spectra in the low and medium luminosity groups, and remains dominant at the hard end of the spectrum in the high luminosity group.  However, its temperature is demonstrably cooler as the luminosity increases, with $kT_{\rm e} > 2$ keV in the low and medium luminosity groups, but $kT_{\rm e} < 2$ keV in the high luminosity group (significant at the $> 90\%$ level in most cases), consistent with the trend noted qualitatively for the stacked \swift spectra.

\subsection{Broadband spectral analysis}
\label{sec:xmm_nustar spectral analysis}

We have shown that the spectral evolution in the 0.3-10 keV range observed by \swift and \xmm appears to follow a set pattern, albeit with some level of spectral degeneracy with luminosity.  However, with \nustar we now have data that extends our bandpass for observing ULXs above 10 keV (e.g. \citealt{bachetti2013}; \citeauthor{walton2013b} 2013b).  Therefore, in this section we extend our analysis to the 0.3-30 keV range\footnote{The \nustar observing bandpass extends to $\sim$79 keV, however no significant detection of \hoix is made above 30 keV \citep{walton2014}.} using two epochs of \xmm and \nustar data taken contemporaneously in 2012 (see Section~\ref{sec:xmm data} and \ref{sec:nustar data}, also \citealt{walton2014}).  We model the \xmm and \nustar spectra from each epoch together, adding a constant multiplicative factor into the model to correct for any calibration differences between all \xmm and \nustar detectors; a $\la$ 10\% difference in the constant parameter is required.  We began by modelling the broadband spectra using the same MCD plus \comptt model used successfully for the stacked \swift and the \xmm spectra.  However, the data reject the model for both epochs of spectral data (null hypothesis probability $\la 0.01$; see Table~\ref{tab:xmm+nustar_fitting_result}).

    \begin{table*}
      \centering
      \caption{The results of joint \xmm and \nustar spectral fits}\label{tab:xmm+nustar_fitting_result}
   
      \smallskip
      \begin{threeparttable}
          \begin{tabular}{lccccccc}

\hline
Model & $N_{\rm H}$$^{a}$ & $kT_{\rm in}$$^{b}$ & $kT_{\rm 0}$ or $\Gamma$ $^{c}$& $\%_{scat}$ $^{d}$&$kT_{2}$ $^{e}$ & $\tau$ or $p$ $^{f}$& $\chi^{2}$/d.o.f. $^{g}$  \\

  & & (keV) & & & (keV) & & \\

\hline             
 \multicolumn{8}{c}{Epoch 1: low/medium luminosity ($f_{\rm X}$ = 1.08 $\times$ 10$^{-11}$ erg cm$^{-2}$ s$^{-1}$)$^{h}$}\\ 
{\sc diskbb+comptt} 	&$	0.22	_{	-0.02	}^{+	0.04	}$&$	0.09	_{	-0.04	}^{+	0.02	}$& -- & -- &$	3.22	\pm0.09	$&$	6.02	\pm0.16	$&$	(1625.53	/	1499)	$\\

{\sc diskbb+(simpl$\times$comptt)}	&$	0.14	\pm0.02	$&$	0.29	_{	-0.02	}^{+	0.06	}$&$	3.80	_{	-0.10	}^{+	0.31	}$&$		>	70			$&$	2.08	_{	-0.12	}^{+	0.08	}$&$	7.47	\pm0.26	$&$	1558.54	/	1497	$\\
{\sc diskbb+(simpl$\times$diskpbb)}	&$	0.16	_{	-0.03	}^{+	0.02	}$&$	0.27	_{	-0.06	}^{+	0.08	}$&$	3.28	_{	-1.35	}^{+	1.38	}$&$	27	_{	-22	}^{+	47	}$&$	3.46	_{	-0.42	}^{+	0.35	}$&$	0.56	_{	-0.01	}^{+	0.01	}$&$	1557.04	/	1497	$\\
																						\hline	
 \multicolumn{8}{c}{Epoch 2: high luminosity ($f_{\rm X}$ = 1.72 $\times$ 10$^{-11}$ erg cm$^{-2}$ s$^{-1}$)$^{h}$} \\															
{\sc diskbb+comptt} 	&$	0.08	\pm0.004$&$	1.20	_{	-0.06	}^{+	0.07	}$& -- & -- &$	2.85	_{	-0.11	}^{+	0.12	}$&$	6.61	_{	-0.53	}^{+	0.75	}$&$	(2082.89	/	1815)	$\\

{\sc diskbb+(simpl$\times$comptt)}$^{gb}$	&$	0.15	_{	-0.01	}^{+	0.02	}$&$	0.32	_{	-0.04	}^{+	0.07	}$&$	3.59	_{	-0.02	}^{+	0.08	}$&$		>	85			$&$	1.06	_{	-0.04	}^{+	0.02	}$&$	11.67	_{	-0.61	}^{+	0.06	}$&$	1869.45	/	1813	$\\
{\sc diskbb+(simpl$\times$comptt)}$^{lc}$ &$	0.12	\pm 0.003$&$	0.58^*					$&$	3.56	_{	-0.02	}^{+	0.08	}$&$		>	81			$&$	1.01	_{	-0.04	}^{+	0.01	}$&$	11.92	_{	-0.02	}^{+	0.08	}$&$	1886.80	/	1813	$\\
{\sc diskbb+(simpl$\times$diskpbb)}	&$	0.14	_{	-0.01	}^{+	0.02	}$&$	1.63	\pm0.06	$&$		>	3.16			$&$	60	_{	-39	}^{+	7	}$&$	3.19	_{	-0.17	}^{+	0.72	}$&$	0.55	\pm	0.01	$&$	1873.24	/	1813	$\\          
\hline             
             
         \end{tabular}
         \begin{tablenotes}
         \item \textbf{Note.} $^{a}$Absorption column beyond our Galaxy, in units of 10$^{22}$ cm$^{-2}$. $^{b}$The inner disc temperature of the MCD component. $^{c}$The seed photon temperature of the \comptt component (in units of keV) or the power-law photon index of the {\sc simpl} component. $^{d}$The scattered fraction of the {\sc simpl} component. $^{e}$The plasma temperature of the \comptt component or the inner disc temperature of the {\sc diskpbb} component. $^{f}$The optical depth of the \comptt component or the value of the $p$ parameter in the {\sc diskpbb} component. $^{g}$Minimum $\chi^{2}$ over degrees of freedom from the fit. $^{h}$Observed X-ray flux in the 0.3-10 keV energy band calculated from the {\sc diskbb+(simpl$\times$comptt)} model. $^{gb}$The best fitting result obtained from the global minimum $\chi^2$ statistic.  $^{lc}$The fitting result obtained from the local minimum statistic (see text).  $^*$We were unable to place constraints on this parameter.         \end{tablenotes}
      \end{threeparttable}
    \end{table*}

It is no surprise that a simple MCD plus \comptt model does not give a good fit, given that this has been demonstrated in several \nustar papers on different ULXs (e.g. \citealt{bachetti2013,walton2014} etc.).  In these papers it is demonstrated that an additional hard component may be present in the data above 10 keV.  We have attempted to account for the hard excess in the data by adding an additional component into the model, a {\sc simpl} model that is an empirical approximation of Comptonisation in which an input seed spectrum is scattered into a power-law component \citep{steiner2009}.  In this we directly follow the models used in \citet{walton2014}, who model the same broadband dataset and propose two different spectral evolution scenarios.  These are based on two models: {\sc diskbb+simpl$\times$diskpbb} and {\sc diskbb+simpl$\times$comptt}.  In the first case the hard component is modelled by an approximation for an advection-dominated `slim' accretion disc in which the temperature profile of the disc is modelled as $T(r) \propto r^{-p}$ and values of $p \approx 0.5$ indicate a slim disc (e.g. \citealt{vierdayanti2006} and references therein); in the second it is modelled by the same \comptt we have been using previously.  

\begin{figure*}
\begin{center}

\includegraphics[width=8.5cm]{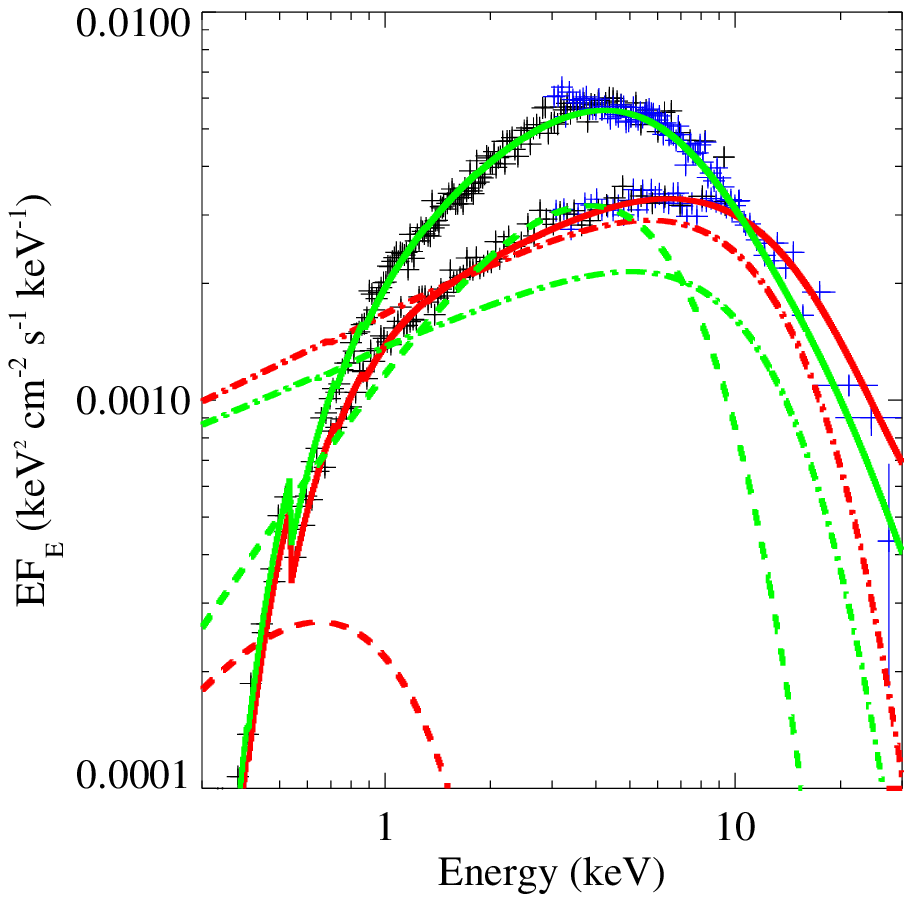}~\includegraphics[width=8.5cm]{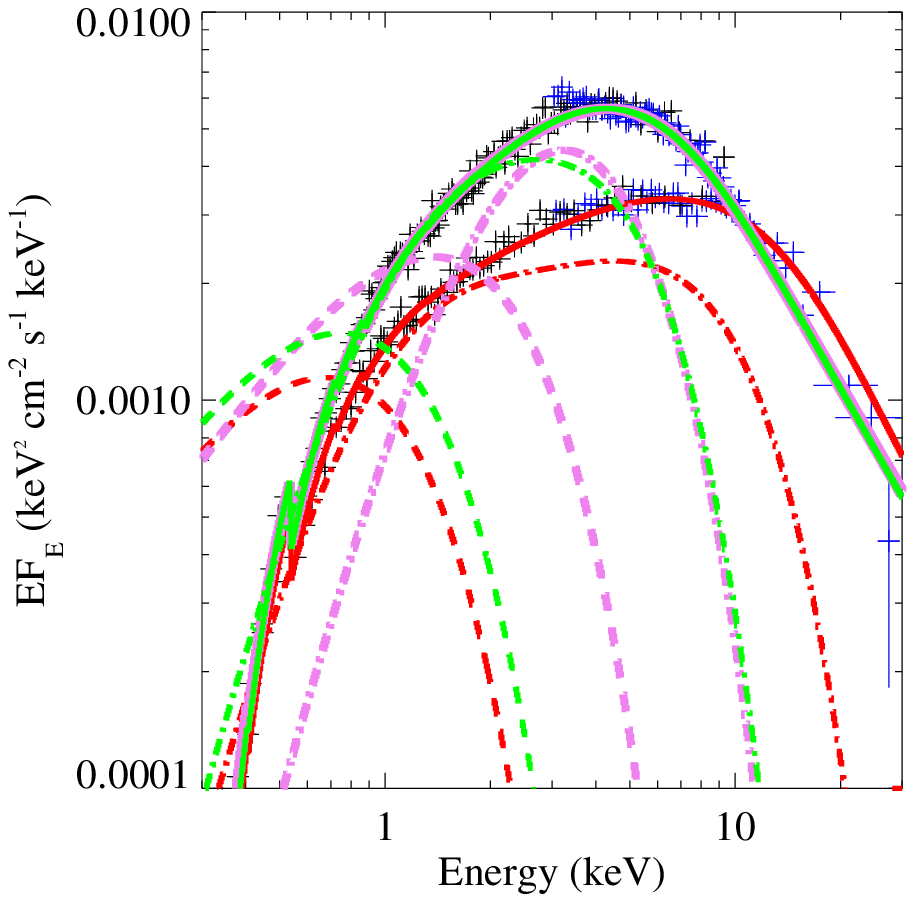}

\caption[]{The {\sc diskbb+simpl$\times$diskpbb} model ({\it left panel}) and {\sc diskbb+simpl$\times$comptt} model ({\it right panel}), fitted to the broadband spectra of \hoix observed by \xmm (black) and \nustar (blue).  The best fitting models are plotted as solid lines for both epoch 1 (red, lower flux) and epoch 2 (green) spectra.  The MCD components, corrected for absorption, are plotted as dashed lines, whilst the {\sc diskpbb} or  {\sc comptt} components, corrected for absorption and modified by the {\sc simpl} model, are plotted as the dash-dotted lines.  The violet lines in the right panel are the fitting result for epoch 2 spectra obtained from the local model minimum (see text).}
\label{fig:xmmnustar_modelplot}
\end{center}
\end{figure*}

Both models fit equally well to the data and the spectral evolution of the models with luminosity is similar to that is described in \citet{walton2014}.  In brief, the evolution obtained from {\sc diskbb+simpl$\times$diskpbb} suggests that evolution is mainly in the MCD component, that increases dramatically in both temperature (from $\sim 0.3$ to $\sim 1.6$ keV) and flux ($\sim 4$ to $\sim 44\%$ of the total unabsorbed flux) in the high luminosity epoch, whilst the {\sc diskpbb} component appears much closer to constant between the two epochs (see the left panel of Fig.~\ref{fig:xmmnustar_modelplot}).   In contrast, the evolution obtained from {\sc diskbb+simpl$\times$comptt} appears in the opposite sense.  The MCD component seems to remain constant whilst the \comptt component appears to play a much larger role in the evolution of the spectra when the luminosity increases (although in both cases very high scattered fractions for the {\sc simpl} component are required, $> 70$ per cent), as is shown in the right panel of Fig.~\ref{fig:xmmnustar_modelplot}.  However, we note that the spectral evolution obtained from the {\sc diskbb+simpl$\times$comptt} model may not be unique; in fact, we found a local minimum in the $\chi^{2}$ fitting of epoch 2 spectra where the $\Delta\chi^{2}$ between the global and local minimum is only $\sim$20 over 1813 degrees of freedom, so we cannot reject the result obtained from this minimum.  Interestingly, the spectral evolution suggested by the local minimum is more consistent with the evolution obtained from the MCD plus \comptt model, particularly for the \swift and \xmm data, in which the MCD component gets stronger and warmer, and the \comptt component peaks at lower energy when the luminosity increases. We will discuss this further in Section~\ref{sec:discussion}.

\section{Discussion} \label{sec:discussion}

\hoix is a persistently luminous ULX, with a decade of monitoring by \swift showing a dynamical range of a factor $\la 5$ and a typical luminosity of around $\sim 10^{40} \rm ~erg~s^{-1}$.  In this paper we have examined the spectral evolution of this ULX with changes in luminosity, based on stacking \swift spectra in terms of their flux and as a combination of their flux and photon index (with the latter as a proxy for X-ray colour), on \xmm pointed observations, and on combining contemporaneous \xmm and \nustar data to provide a broader bandpass in which to study the variation of the ULX.  Our main finding is that there appears to be a common trend in the changing morphology of the spectrum as the luminosity increases, albeit with evidence for some level of degeneracy with luminosity.  At lower luminosities (around $10^{40} \rm ~erg~s^{-1}$) we see the spectra as either a distinct, two component ULX spectrum with a dominant hard component, or as a hard, flat spectrum in the 1-6 keV range, with a turnover in the spectrum at higher energies.  These spectra would classify \hoix as in the {\it hard ultraluminous\/} regime of \citet{sutton2013}.  However, as the luminosity increases to $\sim 2 \times 10^{40} \rm ~erg~s^{-1}$, its shape changes to become much more curved and disc-like, with a single peak at energies of $\sim$ 4-5 keV (indeed, it is likely that some of these spectra would be classified as {\it broadened disc} according to the \citealt{sutton2013} scheme).

The spectra in the 0.3-10 keV bandpass of \swift and \xmm are all satisfactorily described by a model composed of two thermal components, a multi-colour disc blackbody model for the soft end of the spectrum, and an optically thick, cool Comptonising corona for the hard end, with the seed photon temperature of the latter fixed at the temperature of the disc.  Such a model has been found to be a good empirical description of ULX spectra in this bandpass, albeit with strong caveats against the direct physical interpretation of this model (e.g. \citealt{gladstone2009}), and allows us to track the variation of the two components that make up the model.  However, when the spectral bandpass is extended above 10 keV by the use of \nustar data, we find that this model no longer constitutes an acceptable fit to the data, as has been found for multiple \nustar observations of ULXs where additional hard flux is required (although this is somewhat model dependent in several ULXs; \citealt{walton2015} and references therein). Here, we accounted for the extra hard flux using an additional component, the {\sc simpl} model.  Two models -- {\sc diskbb+simpl$\times$diskpbb} and {\sc diskbb+simpl$\times$comptt} -- were used to fit the data, and they were found to equally well describe the broadband spectra.

\subsection{The model-dependent variability of individual spectral components}

In this work, we primarily used a MCD plus Comptonisation model to analyse the ULX spectra.  Although this model has strong caveats about its direct physical interpretation (see above), tracking the variability of each individual spectral component might provide some useful information about the change in the properties and/or geometry of the accretion disc.  We found that the behaviour of the individual MCD and \comptt components appear to follow general patterns in the 0.3-10 keV data, although this is somewhat confused by the degenerate behaviours.  In most of the \swift and \xmm data the variability is mainly in the Comptonised component, that brightens but appears to drop to a lower peak energy (i.e. cools down) as the total luminosity increases.  The MCD component is mostly relatively stable, although it does appear to brighten and warm up in the most luminous datasets.  However, the behaviour becomes much less certain when the \nustar data is included. Here, we tried multiple models in common with other work, and found that the variability of individual components seems to be dependent on the model used to fit the spectra.  In the modified version of the MCD plus Comtonisation model, with an additional {\sc simpl} component to represent the required hard excess, the evolution is uncertain, with the \comptt cooling but fits with the MCD both remaining the same, and increasing in temperature and flux, are permitted.  The situation is confused further by the use of a {\sc diskpbb} `slim disc' model in place of the \comptt, in which case the variability is dominated by the MCD and the slim disc remains relatively stable between the two epochs. Thus, the real evolution of the soft and hard components appears highly uncertain, with the evolution we see depending upon the models we use to fit the data.

\begin{figure*}
\begin{center}

\includegraphics[width=8.5cm]{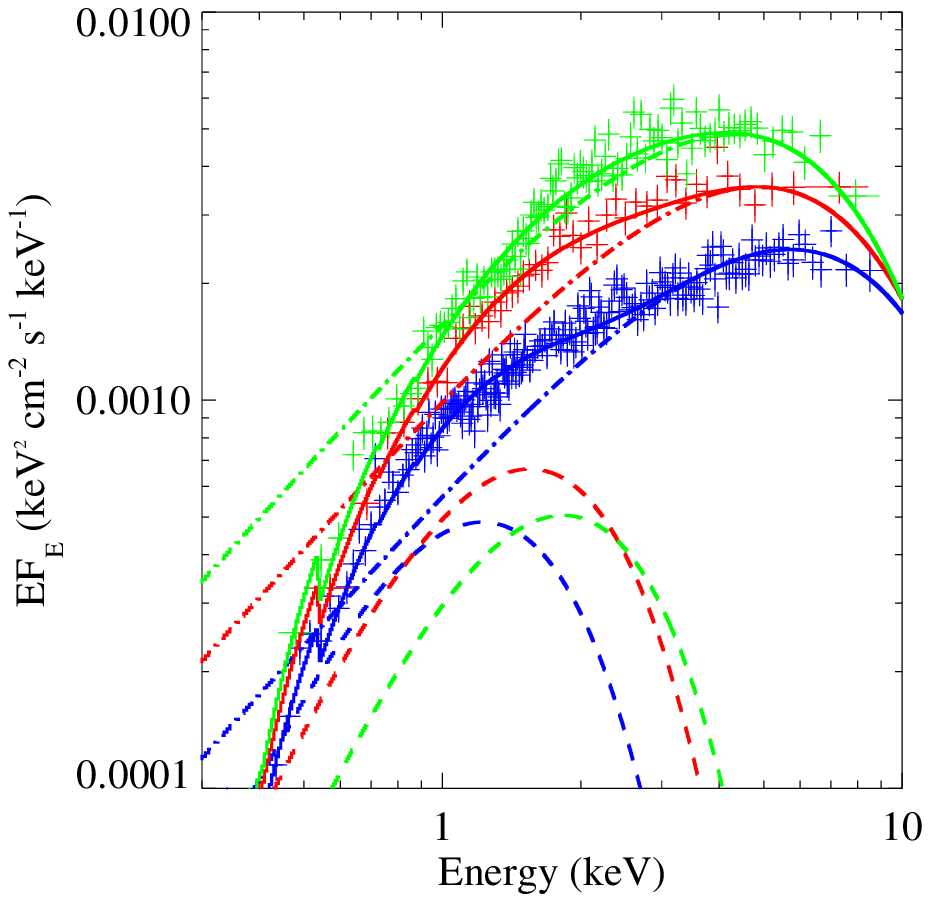}~\includegraphics[width=8.5cm]{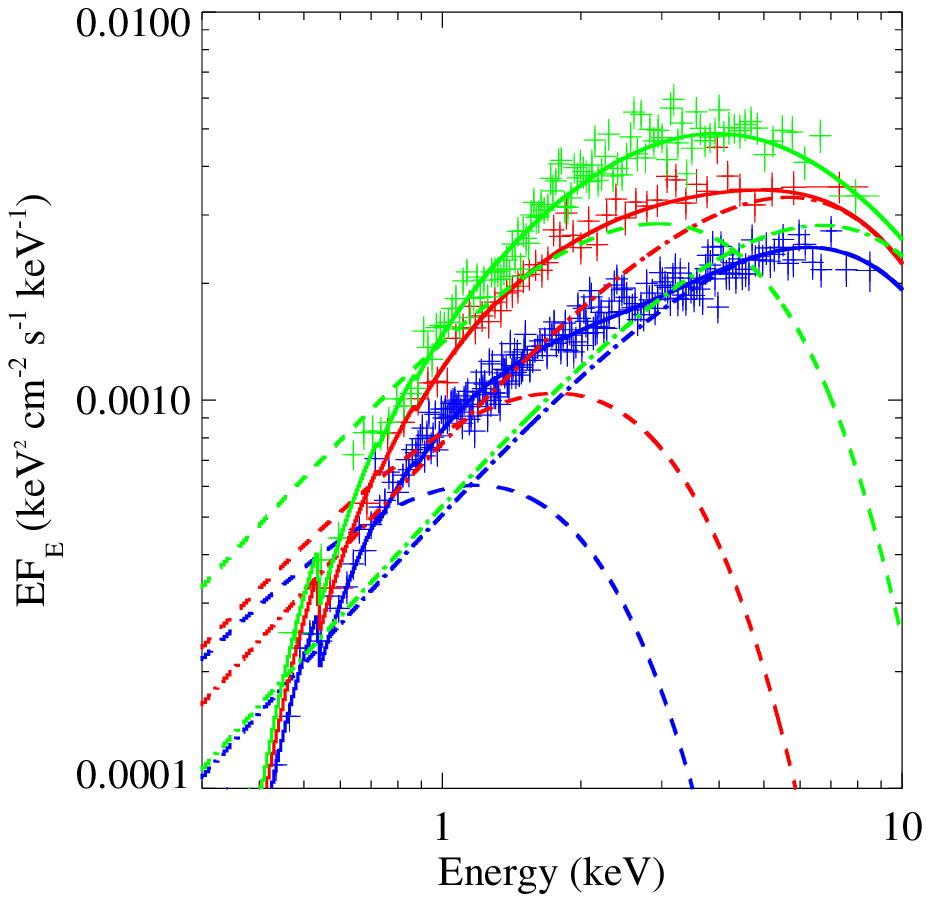}

\caption[]{The \swift luminosity bin spectra fitted by the  {\sc bb} + {\sc diskbb} model ({\it left panel}) and {\sc diskbb} + {\sc diskbb} model ({\it right panel}). Only on-axis spectra are shown and they are re-binned to a minimum of 10$\sigma$ statistical significance per data point to clarify the plot.  The line styles, colours and order of the spectra are as per Fig.~\ref{fig:swift_lum_bin_spectra}, excepting for the dashed lines and the dash-dotted lines which here represent the lower energy components ({\sc bb} for the {\it left panel}  and lower temperature {\sc diskbb} for the {\it right panel}) and higher energy components ({\sc diskbb} for the {\it left panel} and higher temperature {\sc diskbb} for the {\it right panel}), respectively.}
\label{fig:compare_model_variability}
\end{center}
\end{figure*}

If this is the case when we fit the broad-band data, then this raises real questions over the fits in the more limited 0.3-10 keV band.  Specifically, is the behaviour of the two components similarly model-dependent in this band?  We investigate this by re-modelling the \swift luminosity bin spectra using two alternative simple, empirical models (with, particularly in the latter case, no direct physical interpretation for a ULX) that have been shown to fit ULX data in this band \citep{stobbart2006}, namely a blackbody plus MCD model ({\sc bb} + {\sc diskbb} in {\sc xspec}) and a double MCD component model ({\sc diskbb} + {\sc diskbb} in {\sc xspec}).   The fitting results show that both models can adequately describe the spectra, with fits that are nearly or equally as good as the {\sc diskbb} + \comptt model in Section~\ref{sec:lum_bin_analysis} (the blackbody plus MCD fit is generally worse by $\Delta \chi^2$ of $\sim 30$ for the \xmm spectra, but still statistically acceptable).  We show these fits, and their underlying component variability, in Fig.~\ref{fig:compare_model_variability}.   Comparing the variability of the individual components seen in these two models with that in the MCD plus Comptonisation model (c.f. Fig.~\ref{fig:swift_lum_bin_spectra}), we see clear differences.  Whilst the variability of the high energy component seems to follow a similar pattern across all three models -- lowering in peak energy as the total luminosity increases -- the low energy components exhibit very different behaviours, in terms of both their flux and temperature.  In the new models, the blackbody appears relatively stable and contributes little to the spectrum, whereas in the dual MCD model the cooler MCD brightens and heats up substantially as the ULX luminosity increases.  Both these behaviours for the soft component are distinct from that seen in the MCD plus \comptt model.  Thus, the behaviour of the two components in the narrower 0.3-10 keV range also appears highly model-dependent.

Independent of the spectral fits, an examination of the data shows that a remarkable feature of all the spectra is that {\it the data appear to show relatively little dynamical range in flux below 1 keV\/}, particularly compared to the 1-6 keV regime.  In fact, if we examine the range in spectral normalisation at both 0.5 keV and 3 keV in Figs.~\ref{fig:swift_lum_bin_spectra}~\&~\ref{fig:xmm_spectra}, we see that the 0.5 keV normalisation varies by factors $1.8  - 2$ in the two figures, compared to a variation of $4 - 4.3$ times at 3 keV.  Hence the variation in relative normalisation is a factor $2 - 2.5$ smaller at the lower energy.  This relative lack of variability in the lower energy range is strongly suggestive that any soft component in the spectrum is not varying strongly.  Therefore, we can infer that the strong changes we do see in the modelled soft component may primarily be driven by the changes in the 1-6 keV band, with a hotter and brighter MCD required to help model the strongly curved spectrum seen at the highest luminosities.  We therefore strongly suspect that some of the changes in the parameterisation of the two component models may be due to the limitations of the models in describing the changing shape of the spectrum, rather than representative of real underlying physical changes.  Hence, we limit the remainder of our discussion to the implications of the change in the overall shape of the spectrum with luminosity for the underlying source physics, and do not discuss the two separate spectral components hereafter, except for one final point.

The behaviour of the soft component in ULX spectra has been a point of controversy, with various studies claiming that its temperature increases with source luminosity, as would be expected for an accretion disc (e.g. \citealt{miller2004b,miller2013}), and others claiming that it decreases with luminosity, as would be expected for an optically thick outflow (e.g. \citealt{kajava2009}).  We note here that our analysis clearly shows that some models require the soft component to heat up and brighten, similar to the expected behaviour of a disc; others show little change.  However, the implication of the above discussion (particularly the relatively low dynamical range of the flux changes below 1 keV) is that this heating and brightening (where seen) is potentially an artefact of using an incorrect model on the data; it appears that it is the overall change in the spectral shape from a flat/two-component spectrum to a narrower, disc-like spectrum in the 1-6 keV range that is driving the soft component to appear to brighten and heat up.  So, in this ULX at least (that we note featured in previous samples suggesting disc-like behaviour for the soft component), we do not think that the evidence for the soft component behaving like a true accretion disc is reliable.   

For the remainder of the discussion we will focus on what the overall shape of the spectrum, and its variation, mean for the physics of this ULX.  We will start by discussing what the implications of our results are for the two models discussed as viable possibilities for \hoix by \citet{walton2014}, namely a slim accretion disc, and a wind-dominated model.

\subsection{Slim accretion disc model}

We first consider whether we could be observing the emission of an accretion disc.  In a standard accretion disc we would expect to see that the peak temperature of the inner disc would scale with luminosity as $L$ $\propto$ $T^{4}$, with any changes being directly attributable to changes in the mass accretion rate.  However, in \hoix we see the peak in the spectrum, that is characteristic of the temperature of an optically-thick medium, decrease in energy as the luminosity increases.  Interestingly, this could be consistent with the emission from a slim disc.  At super-critical accretion rates we expect the disc to become geometrically thick as its interior becomes supported by radiation pressure from material advecting directly into the black hole (e.g. \citealt{poutanen2007}).  This means that in sources at high inclination angles, a slim disc can become self-obscuring and so if its scale height increases with luminosity it will appear to get softer as more of its harder central regions become self-obscured \citep{vierdayanti2013}.

However, there are problems with this scenario for \hoix.  Firstly, the indications from both its optical spectra (lack of strong radial velocity variability in its broad emission lines; \citealt{roberts2011}) and its X-ray behaviour (diagnosis as in the hard ultraluminous regime; \citealt{sutton2013}) are that it is viewed close to face-on, and not at a high inclination angle, so self-obscuration by the disc may be difficult to achieve in this object.  Secondly, if the mass accretion rate is varying, then we would expect to see this reflected in variations of the soft part of the X-ray spectrum, as this part of the disc should remain visible, with a rising mass accretion rate leading to a rise in the soft X-ray flux that is particularly pronounced below 1 keV (see e.g. Fig. 1 of \citealt{vierdayanti2013}).  However, as noted above, the variations below 1 keV in the data we analyse are not particularly strong between different spectra, and less than those above 1 keV.  Thus we do not regard this model as a particularly strong candidate for explaining the evolution of spectra with luminosity we see in this ULX.

\subsection{The effect of a massive outflowing wind}
\label{sec:launch of an outflowing wind}

It has been proposed that the soft component seen in the spectra of ULXs can be directly attributed to the presence of an optically thick, massive outflowing wind (e.g. \citealt{kajava2009}).  Theoretically, such an outflowing wind should be launched from the disc at supercritical accretion rates \citep{poutanen2007,ohsuga2011,takeuchi2013}. To explain the spectral variability in the context of an outflowing wind, we consider a spectral-timing model in the supercritical regime as recently described by \citet{middleton2015}.  In this, as the accretion disc enters the supercritical regime its scale height $H/R$ (where $H$ is the height of the disc above its central plane at a radius $R$) increases such that $H/R \sim 1$, similarly to the slim disc model (i.e. the disc becomes geometrically-thick).  However, in this model the intense radiation release from the disc drives its own loosely bound outer layers away, in the form of a massive, radiatively-driven wind, such that the accretion is always locally Eddington-limited.  The disc and wind bound a funnel-shaped, low density region along the rotational axis of the black hole, into which the hard radiation of the inner disc can radiate freely, and/or is scattered along by the inner parts of the wind.  The wind itself is an optically thick medium that emits thermally in soft X-rays.  This geometry is summarised in fig. 1 of \citet{middleton2015}  (see also fig. 2 of \citealt{poutanen2007}), and it is inevitable that in this geometry the observed X-ray spectrum is primarily dependent upon viewing angle.  It is also dependent upon accretion rate, as the opening angle of the funnel is predicted to close as the accretion rate increases, thus driving a more massive wind \citep{king2009}.  This should mean, for a fixed line-of-sight, as the accretion rate increases the changes in spectrum should be predictable (c.f. \citealt{middleton2015}): if viewed close to face-on, and the wind remains out of the line-of-sight, the spectrum should get harder as more of the hard flux is scattered over a smaller solid angle, up the line of sight towards the observer.  If, however, the wind enters the line of sight then the spectrum should soften as the wind scatters some fraction of the hard photons out of the line-of-sight.  In both cases the soft X-ray emission from the wind should increase, as the total outflow rate of the wind is proportional to the mass accretion rate \citep{poutanen2007}.

Critically, this model provides an explanation for the comparative lack of soft variability in Holberg IX X-1.  \cite{king2009} notes that the beaming factor $b$ (i.e. the fraction of the sky that the funnel is open to) is related to the observed luminosity as $L \propto b^{-1}$, and is related to the mass accretion rate $\dot{m}$ as $b \propto \dot{m}^{-2}$.  Thus, for the hard emission emanating from the central regions of the disc, we should see its luminosity increase as $L \propto \dot{m}^2$ if we are viewing down the funnel, which is faster than the luminosity of the isotropic wind emission increases ($\propto \dot{m}$).  This relative hardening of the spectra is also discussed by \citeauthor{middleton2015} (2015, e.g. their Eq.~10), and can both qualitatively and quantitatively explain how our spectra appear to vary more above 1 keV than below, including the dynamic ranges (factor $\sim 2$ difference between the faintest and brightest normalisations at 0.5 keV, compared to $\sim 4$ at 3 keV).  Indeed, given that the luminosity of \hoix is dominated by its hard X-ray emission, its dynamic luminosity range of $\sim 4$ in the reported observations implies that the underlying accretion rate probably varies by a factor of $\sim 2$ over the course of the decade of observations, making it a remarkably stable accretor.

If we are viewing the central regions of the critical accretion flow down the funnel, the evolution of the peak in the spectrum we see from Holmberg IX X-1 still requires an explanation.  As the luminosity increases, this peak falls to lower energies.  Interestingly, this spectral evolution is consistent with simulations of the Comptonised spectra of BHs at extreme supercritical accretion rates as described by \citet{kawashima2012}.   They predict that at lower super-Eddington mass accretion rates, the spectra are harder than expected from a pure slim disc spectrum due to photon up-scattering in the shock-heated region near the BH (a mechanism that we note may provide an explanation for the hard excess seen in \nustar spectra).  However, at higher, extreme supercritical accretion rates, the simulated spectra are softer and become curved and disc-like as the outflow funnel opening angle becomes smaller, so decreasing the number of photons that escape without entering the cool dense outflow (the wind) and being Compton down-scattered. 

Thus we suggest that as the majority of hard photons we observe are scattered up the funnel, they lose some fraction of their energy in this scattering process and so the hard spectral component becomes Compton down-scattered.  As the effect of this process will increase in magnitude as the funnel narrows and so a higher fraction of photons become down-scattered, this naturally explains the lower peak temperatures of the hard component as the luminosity of \hoix increases.

\subsection{Source precession}

One feature of the spectra remains to be explained: the spectral degeneracy with luminosity.  This has also previously been reported for this ULX by \citet{vierdayanti2010}, and differences in spectral shape at similar luminosities have also been reported for two other ULXs, IC 342 X-1\citep{marlowe2014} and Ho II X-1 \citep{grise2010}.  These studies agree that the degeneracy demonstrates that the observed spectra of ULXs are not a function of accretion rate alone; \citet{vierdayanti2010} suggest that changes in the structure of the outflowing wind may be culpable for the changes in spectral shape, whereas \citet{marlowe2014} suggest changes in accretion state (possibly between broadened disc and hard ultraluminous regimes in the case of IC 342 X-1).  It is also clear that this degeneracy is not a simple prediction of the wind model -- assuming a fixed line-of-sight, the spectrum will change as described above according to the mass accretion rate, and so we do not expect any strong degeneracy in spectral shape with luminosity except perhaps when the wind comes into the line of sight and starts to diminish the hard flux.  This should, however, only occur above a certain flux threshold, and not at all observed luminosities.  Hence we need a further means of creating the degeneracy; and here we explore one, that the BH rotation axis of \hoix precesses with respect to our line-of-sight, and in doing so induces the degeneracy.

Precession is a known phenomenon of Galactic BHBs, most particularly in the case of SS433 (see \citealt{fabrika2004} for a review).  This is a very apposite example for \hoix as SS433 has been suggested as a hyper-Eddington accretor that would appear as a ULX if viewed face-on.  The spectral variability we would expect from precession can be inferred from \citet{kawashima2012} and \citet{middleton2015}.  Essentially, for a fixed accretion rate, we would expect the spectra to be harder when the line-of-sight is closer to the axis, and softer when further away.  This is again due to the effect of the funnel around the BH rotation axis; closer to on-axis more hard photons from the inner disc are scattered towards the observer, whereas further from axis the edge of the funnel is reached and material in the wind starts to enter the line-of-sight, which both scatters away hard photons (again up the funnel) and Compton down-scatters other hard photons that pass through the lower density regions of the clumpy wind.  Hence, for a fixed accretion rate, we would expect to see a constant soft component, but a hard component that diminishes and softens as the source precesses away from the BH rotation axis.  It is quite plausible that this can explain the differences between the soft and hard spectra we see in Fig.~\ref{fig:swift_hardness-lum_bin_spectra}.  For example, the peak energy in all the soft spectra is lower than in the hard spectra, consistent with down-scattering in the wind.  Additionally, the soft spectra also appear brighter below 2 keV than the equivalent hard spectra; this implies that their accretion rates are likely to be higher, but that hard flux is being lost to scattering in the wind.  A further illustration of this effect is the differences in the MCD component in the different observations in the \xmm high luminosity bin with, for example, observation 0111800101 appearing much softer than the other spectra, with a much higher MCD contribution than two of the four other spectra in this bin despite its lower luminosity.  Finally, interesting supporting evidence for precession might be provided by the long term X-ray variability; it has been recently reported by \citet{lin2015} that Holmberg IX X-1 exhibits a quasi-period of $\sim$ 625 days in {\it Swift} data, potentially a superorbital period due to the precession of the accretion disc.  Hence the predicted changes from this physical process appear largely consistent with the spectral degeneracy that we do see, and are potentially supported by the observed long term X-ray variability, so we infer that precession has an important role in the observed spectrum of this ULX.

\section{Conclusion}
\label{sec:conclusion}

In this paper we have analysed {\it Swift\/}, \xmm and \nustar spectra of the archetypal ULX Holmberg IX X-1.  The wealth of data, particularly in the form of over 500 observations with {\it Swift\/}, has allowed us to study the evolution of the spectra with observed source luminosity.  We find that the data tend to evolve from relatively flat spectra in the 1-6 keV range, or two-component spectra that have a classic hard ultraluminous form at lower luminosities, to a spectrum that is distinctly curved and disc-like at the highest luminosities, with the peak energy in the curved spectrum tending to decrease with increased luminosity.  We study the spectra mainly in terms of a two-component model consisting of a cool multi-colour disc blackbody and a hotter Comptonised medium ({\sc diskbb $+$ comptt} in {\sc xspec}), but the requirement for an additional component in \nustar data, the degeneracy in behaviour of the soft and hard components when other models are used, and the need for the MCD to become substantially brighter and warmer when the disc spectrum gets very curved, dissuades us from directly interpreting the changes in individual model components.  Instead, we discuss the changes in the overall spectral shape in terms of physical models.  We argue that a `slim disc' model can produce the apparent cooling in the spectra at higher luminosities, if it is viewed at high inclination.  However, other studies argue that \hoix is viewed close to face on, and a relative lack of variability below 1 keV, compared to above, is not expected for a pure disc model.  Instead, a super-critical accretion disc with a massive, radiatively driven wind appears likely to be able to explain the main characteristics of the spectral evolution, with the cooling occurring due to Compton down-scattering in the wind as it expands and starts to cross the line-of-sight at the highest accretion rates.  The relatively higher increase in flux above 1 keV is then due to beaming of the hard emission from the inner disc up the `funnel' structure bounded by the disc and the massive, optically-thick wind.  The remaining characteristic of the data -- a degree of degeneracy between different spectra observed at the same luminosity -- can be explained if the ULX does not remain at a fixed inclination to the line-of-sight, but instead the BH rotational axis precesses.

The analysis in this paper adds to the growing body of evidence that many ULXs are likely to be sMBHs accreting at super-Eddington rates, where their observational properties can be explained by the distinct geometry produced by the combination of the large scale height disc, and the massive, outflowing wind.  The supposition that this geometry changes in response to the accretion rate means that our view of this phenomenon is driven by two main factors: the accretion rate, and the inclination of the BH axis to our line-of-sight.  In the case of \hoix we argue that the range of spectra seen, and particularly the degeneracy with luminosity, mean that both these parameters vary in the ULX; we see the ULX precess, as is seen in the Galactic BH SS433.  So, our understanding of the ULX phenomenon continues to grow.  However, many questions remain, not least the details of the exact geometry of the disc and outflowing wind (if this is indeed the correct model for ULXs), and the state of the material in it.  The answers to some at least of these questions may be accessible in the future if a calorimeter detector is successfully flown, the potential high resolution and large collecting area of which would be ideal for detecting narrow emission and/or absorption feature in the ULX winds, if they are present.
 
\section*{Acknowledgments}

We thank the anonymous referee for their comments, that have helped improve this paper.  WL acknowledges financial support in the form of funding for a PhD studentship from the Royal Thai Government. TPR and CD thank STFC for support as part of the consolidated grant ST/L00075X/1. We also would like to thank Matthew Middleton and Dominic Walton for helpful discussions.  This work is in part based on observations obtained with {\it XMM-Newton\/}, an ESA science mission with instruments and contributions directly funded by ESA Member States and NASA.  This research has also made use of data obtained with NuSTAR, a project led by Caltech, funded by NASA and managed by NASA/JPL, and has utilized the NUSTARDAS software package, jointly developed by the ASDC (Italy) and Caltech (USA).

\bibliography{references}
\bibliographystyle{mn2e}

\bsp

\label{lastpage}

\end{document}